\documentclass[11pt]{article}

\usepackage{amsthm,amsmath,bm,amssymb,fullpage,times}

\newcommand{\ZZ}{\mathbb{Z}}
\newcommand{\CC}{\mathbb{C}}
\newcommand{\FF}{\mathbb{F}}

\newcommand{\onemat}{\leavevmode\hbox{\small1\kern-3.8pt\normalsize1}}

\newcommand{\ee}{{\varepsilon}}

\newtheorem{theorem}{Theorem}
\newtheorem{lemma}[theorem]{Lemma}
\newtheorem{corollary}[theorem]{Corollary}
\newtheorem{fact}[theorem]{Fact}

\newcommand{\PSL}{{\rm PSL}}
\newcommand{\SL}{{\rm SL}}
\newcommand{\GL}{{\rm GL}}
\newcommand{\QFT}{{\rm QFT}}
\newcommand{\rank}[1]{{\rm rank}(#1)}
\newcommand{\tr}[0]{{\rm Tr}}

\newcommand{\E}{\mathrm{E}}

\newcommand{\ext}[0]{\overline}
\newcommand{\ind}[0]{\uparrow}
\newcommand{\res}[0]{\downarrow}

\newcommand{\ket}[1]{|#1\rangle}
\newcommand{\bra}[1]{\langle #1|}
\newcommand{\qform}[3]{\left\langle #1\left|#2\right|#3\right\rangle}

\newcommand{\norm}[1]{\left\|{ #1 }\right\|}
\newcommand{\totvar}[1]{\left\|{ #1 }\right\|_1}
\newcommand{\trnorm}[1]{\left\|{ #1 }\right\|_{\mathrm{tr}}}

\newcommand{\hG}{\widehat{G}}

\newcommand{\hbb}{\widehat{\mathbf{b}}}

\newcommand{\cN}{\mathcal{N}}
\newcommand{\cB}{\mathcal{B}}
\newcommand{\cM}{\mathcal{M}}
\newcommand{\cP}{\mathcal{P}}
\newcommand{\cS}{\mathcal{S}}
\newcommand{\cF}{\mathcal{F}}

\newcommand{\tH}{\tilde{H}}
\newcommand{\tG}{\tilde{G}}
\newcommand{\tg}{\tilde{g}}

\newcommand{\zeromat}[0]{{\mathbf 0}}
\newcommand{\bb}{\mathbf{b}}
\newcommand{\bi}{\mathbf{i}}

\newcommand{\bcB}{\boldsymbol{\mathcal{B}}}
\newcommand{\bcS}{\boldsymbol{\mathcal{S}}}
\newcommand{\brho}{\boldsymbol{\rho}}
\newcommand{\bsigma}{\boldsymbol{\sigma}}
\newcommand{\btheta}{\boldsymbol{\theta}}

\begin{document}

\title{Limitations of Quantum Coset States for Graph Isomorphism}

\author{
  Sean Hallgren, Martin R{\"o}tteler, and Pranab Sen\\
  NEC Laboratories America, Inc.\\
  4 Independence Way, Suite 200\\
  Princeton, NJ 08540, U.S.A.\\
  {\tt $\{$hallgren,mroetteler,pranab$\}$@nec-labs.com}
}

\date{}

\maketitle

\begin{abstract}
It has been known for some time that graph isomorphism reduces to
the hidden subgroup problem (HSP).  What is more, most exponential
speedups in quantum computation are obtained by solving instances of
the HSP.  A common feature of the resulting algorithms is the use of
quantum coset states, which encode the hidden subgroup.  An open
question has been how hard it is to use these states to solve graph
isomorphism.  It was recently shown by Moore, Russell, and
Schulman~\cite{MooreRS05} that only an exponentially small amount of
information is available from one, or a pair of coset states.  A
potential source of power to exploit are entangled quantum
measurements that act jointly on many states at once.  We show that
entangled quantum measurements on at least $\Omega(n \log n)$ coset
states are necessary to get useful information for the case of graph
isomorphism, matching an information theoretic upper bound. This may
be viewed as a negative result because highly entangled measurements
seem hard to implement in general.  Our main theorem is very general
and also rules out using joint measurements on few coset states for
some other groups, such as $\GL(n,\FF_{p^m})$ and $G^n$ where $G$ is
finite and satisfies a suitable property.
\end{abstract}

\section{Introduction}

Almost all exponential speedups that have been achieved in quantum
computing are obtained by solving some instances of the Hidden
Subgroup Problem (HSP).  In particular, the problems underlying Shor's
algorithms for factoring and discrete logarithm \cite{Shor:97}, as
well as Simon's problem \cite{Simon:94}, can be naturally generalized
to the HSP: given a function $f : G \to S$ from a group $G$ to a set
$S$ that is constant on left cosets of some subgroup $H \leq G$ and
distinct on different cosets, find a set of generators for $H$.
Ideally, we would like to find $H$ in time polynomial in the input
size, i.\,e. $\log |G|$.  The abelian 
HSP~\cite{Kitaev:95,BH:97,ME:98}, i.\,e., when $G$ is an
abelian group, lies at the heart of efficient quantum algorithms for
important number-theoretic problems like factoring, discrete
logarithm, Pell's equation, unit group of a number field 
etc.~\cite{Shor:97,Hallgren:2002,Hallgren:2005,SV:2005}.

It has been known for some time that graph isomorphism reduces to the
HSP over the symmetric group~\cite{Beals:97, EHK:99a}, a non-abelian
group. While the non-abelian HSP has received much attention as a
result, efficient algorithms are known only for some special classes
of groups~\cite{IMS:2003,FIMSS:2003,MRRS:2004,BCD:2005b}.  On the
other hand, the HSP presents a systematic way to try and approach the
graph isomorphism problem, and this approach is rooted in developing a
deeper understanding of how far techniques and tools that have worked
in the abelian case can be applied. To the best of our knowledge,
the only other approach to solve
graph isomorphism on a quantum computer is by creating a 
uniform superposition of all graphs isomorphic to a given graph.
It has been proposed to create this superposition 
via quantum sampling of
Markov chains~\cite{AT:2003}, however, very little is known about
this.

One of the key features of a quantum computer is that it can compute
functions in superposition. This fact alone does not lend itself to
exponential speedups, for instance for unstructured search problems it
merely leads to a polynomial speedup~\cite{grover:96,bbbv:97}.
On the other hand, the quantum
states resulting from HSP instances have far more structure since they
capture some periodicity aspects of the function $f$. Evaluating the
function $f$ in superposition and ignoring the function value results
in a random {\em coset state}. Coset states are quantum states of the
form $\ket{gH} = \frac{1}{\sqrt{|H|}} \sum_{h \in H} \ket{gh}$, in
other words, a coset state is a uniform superposition over the
elements of the left coset $gH$. The challenge in using coset states
lies in the fact that $g$ is a random element of the group, beyond our
control, that is, we only have the mixed state $\sigma^G_H =
\frac{1}{|G|} \sum_{g \in G} \ket{gH}\bra{gH}$ and we have to
determine $H$ from it.  Though it is conceivable that some advantage
can be had by making use of the function values, currently there are
no proposals for using function values in any meaningful way.

How much information can be extracted from coset states? The most
general way to extract classical information from quantum states are
POVMs \cite{Nielsen}. A fixed POVM operates on a fixed number $k$ of
coset states at once. This induces a probability distribution over the
set of classical outcomes associated with the POVM. A potential source
of power with no classical analog is that the distribution induced by
a POVM on $k$ coset states may have significantly more information
than a POVM that acts on just one coset state at a time. In other
words, the resulting distribution when the POVM is applied to $k$
coset states can be far from a product distribution. In this case we
say that the POVM is an entangled measurement. The goal of this paper
is to determine how small $k$ can be made such that a polynomial
amount of information about $H$ can be obtained from a POVM on $k$
coset states. More precisely, we want to know how small $k$ can
be so that there exists a POVM on $k$ coset states
that gives polynomially large total variation distance between
every pair of candidate hidden subgroups. Note that this POVM
can have many classical outcomes, and it may have to be repeated
several times if we want to identify the actual hidden subgroup $H$
with constant probability.

In this paper, we show that for many groups $G$ this number $k$
has to be quite large, sometimes as large as $\Omega(\log |G|)$. 
This matches
the information theoretic upper bound of $O(\log |G|)$ for general
groups~\cite{EHK:99}.
Our result can be viewed as a negative result because highly
entangled measurements seem hard to implement in general. Note that
the time required to perform a generic measurement entangled
across $k$ states increases exponentially with $k$. 

For abelian groups the picture simplifies dramatically. Indeed, in
this case a POVM operating on one coset state (i.\,e., $k=1$) exists
that gives a polynomial amount of information about the hidden
subgroup.  Moreover, this measurement is efficiently implementable
using the quantum Fourier transform over the group.  The Fourier based
approach extends to some non-abelian groups as well, e.\,g., dihedral,
affine and Heisenberg groups, and shows that for these groups there
are measurements on single coset states that give polynomially large
information about the hidden subgroup~\cite{EH:98,MRRS:2004,RRS:2005}.

Other than the general information-theoretic upper bound, only a few
examples of
measurements operating on more than one coset state (i.\,e., $k>1$) 
are known that give a polynomial amount of information about 
the hidden subgroup. Kuperberg~\cite{Kuperberg:2003} gave a
measurement for the dihedral group operating on
$2^{O(\sqrt{\log |G|})}$ coset states that also takes
$2^{O(\sqrt{\log |G|})}$ time to implement. 
Bacon et al.~\cite{BCD:2005b} gave 
an efficiently implementable measurement for the Heisenberg
group operating on two coset states, and similar
efficient measurements for some other groups operating on a
constant number of coset states.

The case of the symmetric group $S_n$ has been much harder to
understand. First it was shown that some restricted measurements
related to the abelian case cannot solve the problem ~\cite{HRT:2003}.
Next the non-abelian aspects of the group were attacked by Grigni et
al.~\cite{GSVV:2004} who showed that for hidden subgroups in $S_n$,
measuring the Fourier transform of a single coset state using random
choices of bases for the representations of $S_n$ gives exponentially
little information.  They left open the question whether a clever
choice of basis for each representation space can indeed give enough
information about the hidden subgroup.  Recently, a breakthrough has
been made by Moore, Russell and Schulman~\cite{MooreRS05} who answered
this question in the negative for $k=1$ by showing that any
measurement on a single coset state of $S_n$ gives exponentially
little information, i.\,e., any algorithm for the HSP
in $S_n$ that measures one coset state at a time requires at 
least $\exp(\Omega(n))$ coset states.
Subsequently, Moore and Russell~\cite{MooreR05} extended this result
by showing that any algorithm that jointly measures two coset states 
at a time requires at least
$\exp(\Omega({\sqrt{n}}/{\log n}))$ coset states.  However, their
techniques fail for algorithms that jointly measure three or more
coset states at a time, and they left the $k \geq 3$ case open.

In this paper, we show that no quantum measurement on $k = o(n \log
n)$ coset states can extract polynomial amount of information about
the hidden subgroup in $S_n$. Thus, any algorithm operating on coset
states that solves the hidden subgroup problem in $S_n$ in polynomial
time has to make joint measurements on $k = \Omega(n \log n)$ coset
states, matching the information theoretic upper bound.  Our results
apply to the hidden subgroups arising out of the reduction from
isomorphism of rigid graphs, and rules out any efficient quantum
algorithm that tries to solve graph isomorphism via the standard
reduction to the HSP in $S_n$ using measurements that act jointly on
less than $n \log n$ coset states at a time.

Our lower bound holds for a more general setting: Given a group $G$,
suppose we want to decide if the hidden subgroup is a conjugate of an
a priori known order two subgroup $H$, or the identity subgroup.  We
show a lower bound on the total number of coset states required by any
algorithm that jointly measures at most $k$ states at a time and that
distinguishes between the above two cases.  Our main theorem uses only
properties of $G$ that can be read off from the values of the
characters at the two elements of $H$.  We also prove a {\em transfer
  lemma} that allows us to transfer lower bounds proved for subgroups
and quotient groups to larger groups.  Using our main theorem and the
transfer lemma, we show lower bounds on the order of entangled
measurements required to efficiently solve the HSP using coset states
in groups $\PSL(2,\FF_{p^m})$, $\GL(n,\FF_{p^m})$, and groups of the
form $G^n$, where $G$ a constant-sized group satisfying a suitable
property, including all groups $(S_m)^n$ where $m\geq 4$ is a
constant.  The case of $(S_4)^n$ is interesting, because there is an
efficient algorithm for the HSP making joint measurements on
$n^{O(1)}$ states using the {\em orbit coset} techniques of
\cite{FIMSS:2003}. However, the orbit coset approach creates coset
states not just for the hidden subgroup $H$, but also for various
subgroups of the form $HN$, where $N \unlhd (S_4)^n$. This example
suggests that one way to design efficient algorithms for the HSP
making highly entangled measurements may be to use coset states for
subgroups of $G$ other than just the hidden subgroup $H$.

Recently, Childs and Wocjan~\cite{CW:2005} proposed a {\em hidden
  shift} approach to graph isomorphism. They established a
lower bound for the total number of hidden shift states required and
also showed that a single hidden shift state contains
exponentially little information about the isomorphism.  Our results
generalize both their bounds and imply that $o(n \log n)$ hidden
shift states contain exponentially little information about the
isomorphism.

The chief technical innovation required to prove our main theorem is
an improved upper bound for the second moment of the probability of
observing a particular measurement outcome as we vary over different
candidate hidden subgroups.  In particular, we give a new and improved
analysis of the projection lengths of vectors of the form $\bb \otimes
\bb$ onto homogeneous spaces of irreducible representations of a
group.  The earlier works~\cite{MooreRS05, MooreR05} tried to bound
these projection lengths using simple geometric methods. As a result,
their methods failed beyond $k=2$ for the symmetric group.  Instead,
we make crucial use of the representation-theoretic structure of the
projection operators as well as the structure of the vectors, in order
to prove upper bounds on the projection lengths better than those
obtainable by mere geometry. This allows us to prove a general theorem
that applies with large $k$ for many groups.

Finally, we also prove a simple lower bound on the total number of
coset states required by any algorithm to solve the HSP in a group
$G$. This lower bound gives a simple proof of the fact that
distinguishing a hidden reflection from the identity subgroup in the
dihedral group $D_n$ requires $\Omega(\log n)$ coset states.

\section{Preliminaries}

\subsection{Graph isomorphism and HSP} 
The usual reduction of deciding isomorphism of two $n$-vertex graphs 
to HSP in
$S_{2n}$ actually embeds the problem into a proper
subgroup of $S_{2n}$, namely, $S_n \wr S_2$~\cite{EHK:99a}. 
The elements of $S_n \wr S_2$ are tuples
of the form $(\pi, \sigma, b)$ where $\pi, \sigma \in S_n$ and $b \in
\ZZ_2$ with the multiplication rule $(\pi_1, \sigma_1, 0) \cdot
(\pi_2, \sigma_2, b) := (\pi_1 \pi_2, \sigma_1 \sigma_2, b)$ and
$(\pi_1, \sigma_1, 1) \cdot (\pi_2, \sigma_2, b) := (\pi_1 \sigma_2,
\sigma_1 \pi_2, 1 \oplus b)$. The embedding of $S_n \wr S_2$ in
$S_{2n}$ treats $\{1, \ldots, 2n\}$ as a union of 
$\{1, \ldots, n\} \cup \{n+1, \ldots, 2n\}$ with 
$\pi, \sigma$ permuting the first and second sets
respectively when $b = 0$, and $\pi$ permuting the first set onto the
second and and $\sigma$ permuting the second set onto the first when
$b = 1$.  There is an element of the form $(\pi, \pi^{-1}, 1)$, called
an {\em involutive swap}, in the hidden subgroup iff the two graphs
are isomorphic.

Additionally, if the two graphs are rigid, i.\,e., have no non-trivial
automorphisms, then the hidden subgroup is trivial if they are
non-isomorphic, and is generated by $(\pi, \pi^{-1}, 1)$ if they are
isomorphic where $\pi$ is the unique isomorphism from the first graph
onto the second. This element $(\pi, \pi^{-1}, 1)$ is of order two,
and is a conjugate in $S_n \wr S_2$ of $h := (e, e, 1)$ where $e \in
S_n$ is the identity permutation.  Viewed as an element of $S_{2n}$,
$h = (1, n+1) (2, n+2) \cdots (n, 2n)$.  The set of conjugates of $h$
in $S_n \wr S_2$ is the set of all involutive swaps $(\pi, \pi^{-1},
1)$, $\pi \in S_n$, and corresponds exactly to all the isomorphisms
possible between the two graphs.  Also $S_n \wr S_2$ is the smallest
group containing all involutive swaps as a single conjugacy class.
This algebraic property makes $S_n \wr S_2$ ideal for the study
of isomorphism of rigid graphs as a hidden subgroup problem.  Note
that graph automorphism, i.\,e., deciding if a given graph has a
non-trivial automorphism, is Turing equivalent classically to
isomorphism of rigid graphs~\cite{KST:93}.

In this paper, we consider the following problem: Given that the
hidden subgroup in $S_n \wr S_2$ is either generated by an involutive
swap or is trivial, decide which case is true.  Graph automorphism as
well as rigid-graph isomorphism reduces to this problem. We show that
any efficient algorithm using coset states that solves this problem
needs to make measurements entangled across $\Omega(n \log n)$ states
(Corollary~\ref{cor:wreath}).  Note that any lower bound for this
problem for a coset state based algorithm holds true even when the
involutive swaps are considered as elements of $S_{2n}$ rather than
$S_n \wr S_2$. This is because of the following general {\em transfer
  lemma}.
\begin{lemma}[Transfer lemma]
\label{lem:transfer}
Let $G$ be a finite group and suppose that either $G \leq \tG$ or 
$G \cong \tG/N$, $N \unlhd \tG$ holds. Then lower bounds for coset 
state based algorithms for the
HSP in $G$ transfer to the same bounds for the HSP in $\tG$ and vice
versa, as long as the hidden subgroups involved are contained in $G$.
\end{lemma}
\begin{proof}
Let $H \leq G$.  The case $G \leq \tG$ follows from the
observation that 
$\CC[\tG] = \bigoplus_{\tg \in \tG/G} L_{\tg} \cdot \CC[G]$,
where $\tG/G$ denotes a system of left coset representatives of $G$ in
$\tG$ and $L_{\tg}$ stands for left multiplication by $\tg$.  Then,
$\sigma^{\tG}_H = \bigoplus_{\tg \in {\tG}/G} L_{\tg} \cdot 
                  \sigma^G_H \cdot L_{\tg}^\dag$, and so any coset 
state based algorithm without loss of
generality performs the same operations on each block of the
orthogonal direct sum.  The case $G \cong \tG/N$ follows from the
observation that $\CC[G]$ is isometric to the subspace of $\CC[\tG]$
spanned by coset states of $N$ namely states of the form 
$\ket{\tg N}$, $\tg  \in \tG$. There is a subgroup 
$\tH \leq \tG$, $N \unlhd \tH$ such that $\tH/N \cong H$. 
Hence, $\sigma^G_H \cong \sigma^{\tG}_{\tH}$.  Thus, any coset
state based algorithm without loss of generality performs the same
operations on $\sigma^G_H$ and $\sigma^{\tG}_{\tH}$.
\end{proof}

Childs and Wocjan~\cite{CW:2005} showed an
$\Omega(n)$ lower bound for the total number of hidden shift states
required to solve graph isomorphism, and also proved that a single
hidden shift state contains exponentially little information about the
isomorphism. However, their results do not rule out 
an algorithm that makes
joint measurements on, say, two states at a time and uses a total of
$O(n)$ hidden shift states. Since 
the hidden shift state corresponding to the shift
$(\pi, \pi^{-1})$, where $\pi \in S_{n/2}$ is exactly the coset state
for the hidden subgroup generated by the involutive swap $(\pi,
\pi^{-1},1)$ in $S_{n/2} \wr S_2$,
Lemma~\ref{lem:transfer} and Corollary~\ref{cor:wreath} of
our paper show that
any efficient algorithm using hidden shift states to solve the graph
isomorphism problem needs to make measurements entangled across
$\Omega(n \log n)$ states, generalizing their results.  

\subsection{Quantum Fourier transform and POVMs}

We collect some standard facts from representation theory of finite
groups; see e.g. the
book by Serre~\cite{Serre:77} for more details.
We use the term irrep
to denote an irreducible unitary representation of a finite group $G$
and denote by $\hG$ a complete set of inequivalent irreps.  For any
unitary representation $\rho$ of $G$, let $\rho^\ast$ denote the
representation obtained by entry-wise conjugating the unitary matrices
$\rho(g)$, where $g \in G$. Note that the definition of $\rho^\ast$
depends upon the choice of the basis used to concretely describe the
matrices $\rho(g)$. If $\rho$ is an irrep of $G$ so is $\rho^\ast$,
but in general $\rho^\ast$ may be inequivalent to $\rho$.  Let
$V_\rho$ denote the vector space of $\rho$, define $d_\rho := \dim
V_\rho$, and notice that $V_\rho = V_{\rho^\ast}$.  The group elements
$\ket{g}$, where $g \in G$ form an orthonormal basis of $\CC^{|G|}$.
Since $\sum_{\rho \in \hG} d_\rho^2 = |G|$, we can consider another
orthonormal basis called the {\em Fourier basis} of $\CC^{|G|}$ 
indexed
by $\ket{\rho, i, j}$, where $\rho \in \hG$ and $i, j$
run over the row and column indices of $\rho$. The quantum Fourier
transform over $G$, $\QFT_G$ is the following linear transformation:
\[
\ket{g} \mapsto \sum_{\rho\in \hG} \sqrt{\frac{d_\rho}{|G|}}
\sum_{i, j = 1}^{d_\rho} \rho_{i j}(g) \ket{\rho, i, j}.
\]
It follows from Schur's orthogonality relations 
(see e.g.~\cite[Chapter 2, Proposition 4, Corollary 3]{Serre:77})
that $\QFT_G$ is
a unitary transformation in $\CC^{|G|}$.

For a subgroup $H \leq G$ and $\rho \in \hG$, 
define $\rho(H) := \frac{1}{|H|} \sum_{h
  \in H} \rho(h)$. It follows from Schur's lemma 
(see e.g.~\cite[Chapter 2, Proposition 4]{Serre:77})
that $\rho(H)$ is an
orthogonal projection to the subspace of $V_\rho$ consisting of
vectors that are point-wise fixed by every $\rho(h)$, $h \in H$.
Define $r_\rho(H) := \mbox{\rm rank}(\rho(H))$; then $r_\rho(H) =
\frac{1}{|H|} \sum_{h \in H} \chi_\rho(h)$, where $\chi_\rho$ denotes
the character of $\rho$. Notice that $r_\rho(H) = r_{\rho^\ast}(H)$.
For any subset $S \leq G$ we define $\ket{S} := \frac{1}{\sqrt{|S|}}
\sum_{s \in S} \ket{s}$ to be the uniform superposition over the
elements of $S$. The {\em standard method} of attacking the HSP
in $G$ using coset states~\cite{GSVV:2004} starts by
forming the uniform superposition $\frac{1}{\sqrt{|G|}} \sum_{g \in G}
\ket{g}\ket{0}$. It then queries $f$ to get the superposition
$\frac{1}{\sqrt{|G|}} \sum_{g \in G} \ket{g} \ket{f(g)}$. Ignoring the
second register the reduced state on the first register becomes the
density matrix $\sigma^G_H = \frac{1}{|G|} \sum_{g \in G}
\ket{gH}\bra{gH}$, that is the reduced state is a uniform mixture over
all left coset states of $H$ in $G$.  It can be easily seen that
applying $\QFT_G$ to $\sigma^G_H$ gives us the density matrix
$\frac{|H|}{|G|} \bigoplus_{\rho \in \hG} \bigoplus_{i=1}^{d_\rho}
\ket{\rho,i}\bra{\rho,i} \otimes \rho^\ast(H)$, where $\rho^\ast(H)$
operates on the space of column indices of $\rho$. 
When measuring this state, we
obtain an irrep $\rho$ with probability 
$\frac{d_{\rho} |H| r_{\rho(H)}}{|G|}$. 
Conditioned on measuring $\rho$ we obtain a
uniform distribution $1/d_\rho$ on the row indices. The reduced state
on the space of column indices
after having observed an irrep $\rho$ and a row index $i$ is then
given by the state $\rho^\ast(H)/r_\rho(H)$, and a basic task for a
hidden subgroup finding algorithm
is how to extract information about $H$ from it.

If the the hidden subgroup is the trivial subgroup $\{1\}$,
the probability of measuring $\rho$ is given by the so-called
Plancherel distribution $\cP(\rho) := \frac{d_\rho^2}{|G|}$. This
distribution will be useful to us later on in the proof of the main
theorem.

POVMs are the most general way to obtain classical information from
quantum states \cite{Nielsen}. The elements of a POVM $\cM$
in $\CC^n$ are
positive operators $E_i\geq 0$ which have to satisfy the completeness
condition $\sum_i E_i = \onemat_n$. If the state of the quantum system
is given by the density matrix $\sigma$, then the probability $p_i$ to
observe outcome labeled $i$ is given by the Born rule $p_i =
\tr(\sigma E_i)$. The following observation is crucial for the HSP
case: since the states $\sigma_H^G$ are simultaneously
block diagonal in the Fourier
basis for any $H\leq G$, the elements of any POVM $\cM$
operating on these
states can without loss of generality be assumed to have the same
block structure.  From this it is clear that any measurement to
identify $H$ without loss of generality first applies the quantum
Fourier transform $\QFT_G$ to $\sigma_H^G$, measures the name $\rho$
of an irrep, the index $i$ of a row, and then measures the reduced
state on the column space of $\rho$ using a POVM $\cM_\rho$
in $\CC^{d_\rho}$. This POVM $\cM_\rho$ may depend on 
$\rho$ but is independent of $i$.

Furthermore, $\cM_\rho$ can be assumed to be a frame, i.\,e., a
collection $\cB_\rho := \{ (a_b, b) \}$, where $b \in \CC^{d_\rho}$ 
with $\norm{b}=1$ and $0 \leq a_b \leq 1$ such that 
$\sum_{b \in \cB} a_b \ket{b}\bra{b} = \onemat_{d_\rho}$ i.e.
a frame is a POVM with rank one elements.
Orthonormal bases are special cases of frames in which 
$a_b = 1$ for all $b \in \cB_\rho$.
We can assume that the POVM on the column space is a frame because
any POVM can be refined to a frame such that for any quantum state,
the probabilities 
according to the original POVM are certain sums, independent of the
state measured, of probabilities according to the frame.

If the the hidden subgroup is the trivial subgroup $\{1\}$, after
observing an irrep $\rho$ and a row index $i$, the reduced state on
the space of column indices of $\rho$ is the totally mixed state
$\frac{\onemat_{d_\rho}}{d_\rho}$.  The probability of observing a
vector $b$ in frame $\cB_\rho$ is given by the so-called natural
distribution on $\cB_\rho$ defined by $\cN(b \mid \rho) :=
\frac{d_\rho^2}{|G|}$. This distribution will be useful to us later on
in the proof of the main theorem.

The above description was for single register quantum Fourier
sampling. Fourier sampling on $k$ registers can be defined
analogously.  Here one starts off with $k$ independent copies of the
coset state $\sigma_H^G$, i.\,e., with the state
$(\sigma_H^G)^{\otimes k} \cong \sigma^{G^k}_{H^k}$ and applies
$\QFT_G^{\otimes k}$ to it. Here $G^k$, $H^k$ denote the $k$-fold
direct product of $G$, $H$ respectively. Note that since
$\widehat{G^k} \cong \hG^{\otimes k}$, we have that $\QFT_{G^k} =
\QFT_G^{\otimes k}$.  We can express an irrep $\brho$ of $G^k$ as
$\brho = \otimes_{i=1}^k \rho_i$, $\rho_i \in \hG$; observe that
$V_{\brho} = \otimes_{i=1}^k V_{\rho_i}$. We adopt the convention
that multiregister vectors and representations are denoted in
boldface type. After applying
$\QFT_G^{\otimes k}$, we measure the name $\brho$ of an irrep of
$G^k$, i.\,e, irreps $\rho_1, \ldots, \rho_k$ of $G$. After that, we
measure a row index of $\brho$ i.\,e., row indices of $\rho_1, \ldots,
\rho_k$, and then measure the resulting reduced state in the column
space of $\brho$ using a frame $\bcB$ of $V_{\brho}$.  The frame
$\bcB$ used depends on the observed $\brho$ but not on the observed
row indices.  Notice that only the application of the frame $\bcB$ may
be an entangled measurement, the application of $\QFT_{G^k}$ and
measurement of $\brho$ together with a row index of $\brho$ are single
register operations.

\section{The main theorem}

Let $G$ be a group and $h\in G$ be an involution, that is, $H := \{1,
h\}$ is an order two subgroup of $G$.  We let $H^g := g H g^{-1}$
denote the conjugate of $H$ by $g \in G$.  Let $k$ be a positive
integer.  Fix a POVM $\cM$ on $\CC[G]^{\otimes k} \cong \CC[G^k]$.
Let $\cM_{H^g}$, $\cM_{\{1\}}$ denote the classical probability
distributions obtained by measuring the states $\sigma_{H^g}^{\otimes
  k}$, $\sigma_{\{1\}}^{\otimes k}$ respectively according to $\cM$.
We will show that the average total variation distance between
$\cM_{H^g}$ and $\cM_{\{1\}}$ over conjugates $H^g$, $g \in G$ is at
most $2^k$ times a quantity that depends purely on the pair $(G, H)$.
In the next section, we will show that this quantity is exponentially
small for many pairs $(G, H)$ of interest, including when $G = S_n \wr
S_2$ and $H$ is generated by an involutive swap, i.\,e., the case
relevant to isomorphism of rigid graphs.
\begin{theorem}[Main theorem]
\label{mainTheorem}
Let $G$ be a finite group and $H:=\{1,h\}$ be an order two subgroup of
$G$. Let $k\geq 1$ be an integer. Fix a POVM $\cM$ on
$\CC[G]^{\otimes k}$ and let $\cM_{H^g}$, $\cM_{\{1\}}$ denote the
classical probability distributions obtained by measuring the states
$\sigma_{H^g}^{\otimes k}$, $\sigma_{\{1\}}^{\otimes k}$ respectively
according to $\cM$.  For $\ee > 0$, define the set 
\[
\cS_\ee :=
\left\{\tau \in \hG: \frac{|\chi_\tau(h)|}{d_\tau} \geq \ee \right\}.
\]
Suppose that $2 k \ee < 1$ holds.  Define
\[
\delta_1 := 
\ee + \frac{1}{|G|} \cdot
\left(\sum_{\tau \in \cS_\ee} d_\tau |\chi_\tau(h)|\right) \cdot
\left(\sum_{\nu \in \hG} d_\nu\right)
\leq \ee + 
\left(\sum_{\tau \in \cS_\ee} d_\tau |\chi_\tau(h)|\right) 
\cdot
\left(\frac{|\hG|}{|G|}\right)^{1/2},
\]
and
\[
\delta_2 := 
2^k(1 + 2 k \ee) \delta_1^{1/2}
+ 3 k \ee + 
\frac{3 k}{|G|} \cdot \sum_{\tau \in \cS_\ee} d_\tau^2.
\]
Then 
\[
\E_g[\totvar{\cM_{H^g} - \cM_{\{1\}}}] \leq \delta_2, 
\]
where
the expectation is taken over the uniform distribution on $g \in G$.
\end{theorem}

By a $k$-entangled POVM $\cF$ on $t$ coset states, we mean that
$\cF$ consists of a sequence of POVM's $(\cM_i)_{i \in [t']}$, where 
each $\cM_i$ operates on a fresh set of at most $k$-coset states and
$t' \leq t$. The number of coset states operated upon by
$\cF$ is at most $t$. The outcome of $\cF$ is a sequence of length
$t'$ corresponding to the outcomes of $\cM_i$. The choice of
$\cM_i$ may depend on the observed outcomes of 
$\cM_1, \ldots, \cM_{i-1}$. If required, further classical 
postprocessing may be done on the outcome of $\cF$. We now
prove the following corollary of Theorem~\ref{mainTheorem}.
\begin{corollary}
Suppose $\cF$ is a $k$-entangled POVM on $t$ coset states. Then
for at least a fraction of $1 - \sqrt{t \delta_2}$ conjugate 
subgroups
$H^g$, $g \in G$, 
\[
\totvar{\cF_{H^g} - \cF_{\{1\}}} \leq \sqrt{t \delta_2}.
\]
\end{corollary}
\begin{proof}
Using Theorem~\ref{mainTheorem} and triangle inequality, it is easy
to see that
$\E_g[\totvar{\cF_{H^g} - \cF_{\{1\}}}] \leq t \delta_2$.
Applying Markov's inequality to the expectation over 
$g \in G$ finishes the proof.
\end{proof}
The remainder of the section is devoted to proving 
Theorem~\ref{mainTheorem}. We first give some notation that will
be useful for the proofs of various lemmas. Our notation and setup
is inspired to a large extent by the notation in \cite{MooreR05}.

As argued in the previous section, we can assume without loss of
generality that $\cM$ first applies $\QFT_G^{\otimes k}$ to 
$\sigma_{H^g}^{\otimes k}$,
measures the name of an irrep of $G^k$, 
$\brho^\ast$
together with a row index of $\brho^\ast$, and then measures
the resulting reduced state
in the column space of $\brho^\ast$ using a frame $\bcB$ of
$V_{\brho^\ast} = V_{\brho}$. If $\brho = \otimes_{i=1}^k \rho_i$,
$\rho_i \in \hG$, then
$\brho^\ast = \otimes_{i=1}^k \rho_i^\ast$.
The frame $\bcB$ used depends
on the observed $\brho^\ast$ but not on the observed row indices.

Suppose the hidden subgroup is $H^g$ for some $g \in G$.
It is easy to see that the
probability that $\cM$ measures $\brho^\ast$ 
is given by
\[
\cM_{H^g}(\brho^\ast) 
= \frac{d_{\brho^\ast} |H^g|^k \cdot r_{\brho^\ast}((H^g)^k)}{|G|^k}
= \frac{2^k d_{\brho} r_{\brho}(H^k)}{|G|^k}.
\]
Notice that 
$\cM_{H^g}(\brho^\ast) = \cM_{H}(\brho)$.
Let $\bcB = \{a_{\bb}, \bb\}$, where
$0 \leq a_{\bb} \leq 1$ and 
$\sum_{\bb} a_{\bb} \ket{\bb}\bra{\bb} = \onemat_{V_{\brho}}$.
Then the reduced state in the column space of $\brho^\ast$ is
$\frac{\brho((H^g)^k)}{r_{\brho}(H^k)}$, if $r_{\brho} \neq 0$.
Hence, the probability
of observing a particular $\bb$ conditioned on having observed
$\brho^\ast$ is
\[
\cM_{H^g}(\bb \mid \brho^\ast) = 
\frac{a_{\bb} \qform{\bb}{\brho((H^g)^k)}{\bb}}{r_{\brho}(H^k)},
\]
if $r_{\brho}(H^k) \neq 0$, and $0$ otherwise. Similarly, if the
hidden subgroup is the identity subgroup then
\[
\cM_{\{1\}}(\brho^\ast) 
= \frac{d_{\brho}^2}{|G|^k}
= \cP(\brho),
\]
where $\cP(\cdot)$ is the Plancherel distribution on
irreps of $G^k$. Also
\[
\cM_{\{1\}}(\bb \mid \brho^\ast) = 
\frac{a_{\bb}}{d_{\brho}} 
= \cN(\bb \mid \brho^{\ast}),
\]
where $\cN(\cdot \mid \brho^\ast)$ is the 
{\em natural distribution} corresponding to the frame $\bcB$.

For a non-empty subset $I \subseteq [k]$, define 
$\brho^I := (\otimes_{i \in I} \rho_i) \otimes 
            (\otimes_{i' \in [k]\setminus I}\onemat_{d_{\rho_{i'}}})$,
where $\onemat_{d_{\rho_{i'}}}$ denotes the identity representation
of $G$ of degree equal to that of $\rho_{i'}$.
For non-empty subsets $I_1, I_2 \subseteq [k]$, define 
$\brho^{I_1,I_2} := \brho^{I_1} \otimes \brho^{I_2}$.
For a representation
$\btheta = \otimes_{i=1}^n \theta_i$ of $G^n$, $\theta_i$ 
representation of $G$, we use $\btheta(g)$ as a shorthand for 
$\otimes_{i=1}^n \theta_i(g)$.
For an irrep $\tau \in \hG$, we use $a^{\btheta}_\tau$ to denote the
multiplicity of $\tau$ in the Clebsch-Gordan decomposition of 
$\btheta$, i.\,e. the number of times $\tau$ occurs in $\btheta$ when
$\btheta$ is viewed as a representation of $G$ embedded 
as the diagonal
subgroup of $G^n$. We let $\Pi^{\btheta}_\tau$ denote the orthogonal
projection from $V_{\btheta}$
onto the homogeneous component of $\tau$ in the above
decomposition. 
We use the following shorthand for expectations:
$\E_{\brho}[\cdot]$, $\E_{\bb}[\cdot]$ and $\E_g[\cdot]$ denote
expectations over the Plancherel distribution on irreps, natural
distribution on frame vectors and uniform distribution on elements
of $G$ respectively.

We define a function 
$X: \hG^{\otimes k} \times \bcB \times G \rightarrow [-1,1]$ as 
\[
X(\brho, \bb, g) := \qform{\bb}{\brho((H^g)^k)}{\bb} - \frac{1}{2^k},
\]
where $\bcB$ is a frame for $V_{\brho}$.
The importance of $X$ will become clear in Lemma~\ref{lem:Xtotvar}
below, which shows that
$\E_{\brho,\bb,g}[|X(\brho, \bb, g)|]$ is closely related to
the total variation distance between $\cM_{H^g}$ and $\cM_{\{1\}}$.

We start by proving the following lemma, which is similar to
\cite[Lemma 11]{MooreR05}. The lemma gives us a way to express the
second moment of $X$ in terms of projections of `coupled' frame
vectors $\bb \otimes \bb$ onto homogeneous components corresponding to
irreps $\tau \in \hG$.  The advantage of doing this is that we can now
distinguish between `good' irreps, namely those with
$\frac{|\chi_\tau(h)|}{d_\tau}$ small, and `bad' irreps, namely those
where $\frac{|\chi_\tau(h)|}{d_\tau}$ is large. The contribution of
`good' irreps to the second moment of $X$ is small.  This idea of
distinguishing between `good' and `bad' irreps goes back to
\cite{MooreRS05}.
\begin{lemma}
\label{lem:chioverd}
\[
\E_g[X(\brho,\bb,g)^2] = \frac{1}{4^k} \sum_{I_1, I_2 \neq \{\}}
\sum_{\tau \in \hG} \frac{\chi_\tau(h)}{d_\tau}
\norm{\Pi^{\brho^{I_1,I_2}}_\tau (\bb \otimes \bb)}^2.
\]
\end{lemma}
\begin{proof}
Since
\begin{eqnarray*}
X(\brho,\bb,g) 
& = & 
\frac{1}{2^k}
\left(\qform{\bb}{\onemat_{d_{\brho}}}{\bb} +
      \sum_{I \neq \{\}} \qform{\bb}{\brho^I(ghg^{-1})}{\bb}
\right) - \frac{1}{2^k} \\
& = &
\frac{1}{2^k}
\sum_{I \neq \{\}} \qform{\bb}{\brho^I(ghg^{-1})}{\bb},
\end{eqnarray*}
we get
\begin{eqnarray*}
&   &
\!\!\!\!\!
\!\!\!\!\!
\!\!
\E_g[X(\brho,\bb,g)^2] 
=  \E_g
      \left[\frac{1}{4^k} \sum_{I_1,I_2 \neq \{\}} 
      \qform{\bb}{\brho^{I_1}(ghg^{-1})}{\bb} \cdot
      \qform{\bb}{\brho^{I_2}(ghg^{-1})}{\bb}
      \right] \\
& = & \frac{1}{4^k} \!\!\sum_{I_1,I_2 \neq \{\}} 
      \E_g[\qform{\bb \otimes \bb}{\brho^{I_1,I_2}(ghg^{-1})}
                 {\bb \otimes \bb}] \\
& = &  \frac{1}{4^k} \!\!\sum_{I_1,I_2 \neq \{\}} 
      \E_g\left[\qform{\bb \otimes \bb}
                      {\bigoplus_{\tau\in\hG} a^{\brho^{I_1,I_2}}_\tau
                      \tau(ghg^{-1})}
                      {\bb \otimes \bb}
          \right] \\
& = &  \frac{1}{4^k} \sum_{I_1,I_2 \neq \{\}} 
      \qform{\bb \otimes \bb}
            {\bigoplus_{\tau \in \hG} a^{\brho^{I_1,I_2}}_\tau
             \E_g[\tau(ghg^{-1})]}
            {\bb \otimes \bb} \\
& = &  \frac{1}{4^k} \sum_{I_1,I_2 \neq \{\}} 
      \qform{\bb \otimes \bb}
            {\bigoplus_{\tau \in \hG} a^{\brho^{I_1,I_2}}_\tau
             \frac{\chi_\tau(h)}{d_\tau} \onemat_{V_\tau}}
            {\bb \otimes \bb} \\
& = &  \frac{1}{4^k} \sum_{I_1,I_2 \neq \{\}} \sum_{\tau \in \hG}
      \frac{\chi_\tau(h)}{d_\tau} 
      \norm{\Pi^{\brho^{I_1,I_2}}_\tau (\bb \otimes \bb)}^2.
\end{eqnarray*}
The fifth equality above follows from Schur's lemma.
\end{proof}

Lemma~\ref{lem:chioverd}
takes care of the `good' irreps. However for `bad'
irreps $\tau$, we have to do something to bound  
$\norm{\Pi^{\brho^{I_1,I_2}}_\tau (\bb \otimes \bb)}^2$.
The papers \cite{MooreRS05,MooreR05} tried to bound it
using the following simple geometric argument: If $\bcB$ is an 
orthonormal basis for $V_{\brho}$, then 
$\{\bb \otimes \bb\}_{\bb \in \bcB}$ is an orthonormal set
in $V_{\brho} \otimes V_{\brho}$. Hence the expectation, over the
uniform distribution on $\bcB$, of the above quantity is upper bounded
by $\frac{\rank{\Pi^{\brho^{I_1,I_2}}_\tau}}{d_{\brho}}$.
If $\bcB$ is a POVM rather than
an orthonormal basis, a similar argument can be made.
This simple method works for $k = 1, 2$ for the symmetric group,
but fails for $k \geq 3$. This is because 
$\rank{\Pi^{\brho^{I_1,I_2}}_\tau}$ becomes larger than $d_{\brho}$.
The problem with the simple method is that 
$\rank{\Pi^{\brho^{I_1,I_2}}_\tau}$ can be potentially as large as
$d_{\brho}^2$. This is where
we need new ideas as compared to those in \cite{MooreRS05,MooreR05}.
We use the fact that the projection $\Pi^{\brho^{I_1,I_2}}_\tau$ is
not arbitrary, but is rather the projection onto the homogeneous 
component corresponding to an irrep of $G$. There is an explicit
representation-theoretic formula for such a projection operator
(see e.g.~\cite[Chapter 2, Theorem 8]{Serre:77}).
Using this formula allows us to `decouple'
$\Pi^{\brho^{I_1,I_2}}_\tau (\bb \otimes \bb)$ into an expression
involving only $\brho^{I_1}$ and $\bb$, and 
$\brho^{I_2}$ and $\bb$, that is, it allows us to remove the 
tensor product. This `decoupling' gets around the problem that 
the rank of the projector can be larger than $d_{\brho}$ whereas
the size of the basis $\bcB$ is only $d_{\brho}$. It allows us
to apply a standard corollary of Schur's orthogonality relations
and finally bound the length of the projection of $\bb \otimes \bb$
by a small quantity.

We now state a few facts that will be used in our `decoupling'
arguments.
The next fact is easy to show and was used in the simple geometric
approach of \cite{MooreRS05,MooreR05} to bound
$\norm{\Pi^{\brho^{I_1,I_2}}_\tau (\bb \otimes \bb)}^2$.
\begin{fact}
\label{expected-projection-length}
Let $W$ be a subspace of $V$.  Let 
$\mathcal{B} := \{a_b, b\}$ be a frame for 
$V$. Let $\Pi^V_W$ denote the orthogonal projection from $V$ onto
$W$. Then 
\[\E_b[\norm{\Pi_W^V (b)}^2] = \frac{\dim W}{\dim V},
\]
where the expectation is taken over the natural distribution on 
$\mathcal{B}$.
\end{fact}
The following fact is a special case of \cite[Lemma 12]{MooreR05}, and
can be easily proved by considering the regular representation of
$G^n$.
\begin{fact}
\label{expected-multiplicity}
Let $\btheta := \left(\bigotimes_{i=1}^n \theta_i\right) \otimes 
                \left(\bigotimes_{i'=1}^{n'} \onemat_{d_{i'}}\right)$
be a representation of $G^{n+n'}$,
where $\theta_i \in \hG$ and $\onemat_{d_{i'}}$ is the identity 
representation of $G$ of dimension $d_{i'}$. 
Suppose each $\theta_i$ is chosen independently 
from the Plancherel distribution on $\hG$.  
Fix $\tau \in \hG$.  Let $a_\tau^{\btheta}$ denote
the multiplicity of $\tau$ in the Clebsch-Gordan decomposition of
$\btheta$ i.\,e. viewing $\btheta$ as a representation of $G$ 
embedded as the diagonal subgroup of $G^{n+n'}$.
Then 
\[
\E_{\btheta} \left[\frac{a_\tau^{\btheta}}{d_{\btheta}}\right]
 = \frac{d_\tau}{|G|}.
\]
\end{fact}
The following fact is a standard result in representation theory
(see e.g.~\cite[Chapter 2, Proposition 4, Corollary 3]{Serre:77}), 
and follows from Schur's orthogonality relations.
\begin{fact}
\label{likemub}
Suppose $\tau \in \hG$ and $b \in V_\tau$, $\norm{b} = 1$. Then,
\[
\E_g[|\qform{b}{\tau(g)}{b}|^2] = \frac{1}{d_\tau}.
\]
\end{fact}

We start off the `decoupling' process by the following lemma.
\begin{lemma}
\label{lem:proj-hom}
Fix $I_1,I_2 \subseteq [k]$, $I_1, I_2 \neq \{\}$, 
$\brho \in \hG^{\otimes k}$, 
$\tau \in \hG$ and $\bb \in V_{\brho}$. Then,
\[
\norm{\Pi^{\brho^{I_1,I_2}}_\tau (\bb \otimes \bb)}^2 \leq
\frac{d_\tau^2}{2}
(
\E_g[|\qform{\bb}{\brho^{I_1}(g)}{\bb}|^2] +
\E_g[|\qform{\bb}{\brho^{I_2}(g)}{\bb}|^2].
\]
\end{lemma}
\begin{proof}
\begin{eqnarray*}
\norm{\Pi^{\brho^{I_1,I_2}}_\tau (\bb \otimes \bb)}^2 
&   =  & |\qform{\bb \otimes \bb}
                {\Pi^{\brho^{I_1,I_2}}_\tau}
                {\bb \otimes \bb}| \\
&   =  & |\qform{\bb \otimes \bb}
                {d_\tau 
                 \E_g[\chi_\tau(g)^\ast \brho^{I_1}(g) \otimes
                                        \brho^{I_2}(g)]
                }
                {\bb \otimes \bb}| \\
&   =  & d_\tau
         |\E_g[\chi_\tau(g)^\ast \qform{\bb}{\brho^{I_1}(g)}{\bb} 
               \cdot \qform{\bb}{\brho^{I_2}(g)}{\bb}]| \\
& \leq & d_\tau^2 
         \E_g[|\qform{\bb}{\brho^{I_1}(g)}{\bb}| \cdot
              |\qform{\bb}{\brho^{I_2}(g)}{\bb}|
             ] \\
& \leq & \frac{d_\tau^2}{2} 
         \left(
         \E_g[|\qform{\bb}{\brho^{I_1}(g)}{\bb}|^2] +
         \E_g[|\qform{\bb}{\brho^{I_2}(g)}{\bb}|^2]
         \right).
\end{eqnarray*}
The second equality follows from a standard result in representation
theory describing the projection operator onto a homogeneous 
component corresponding to an irrep of $G$ 
(see e.g.~\cite[Chapter 2, Theorem 8]{Serre:77}), the first inequality
follows by bounding a character value by the dimension of the
representation, and the second 
inequality
follows from the fact that $|xy| \leq \frac{|x|^2 + |y|^2}{2}$ for any
pair of complex numbers $x,y$.
\end{proof}

We now prove a crucial lemma that allows us to
prove good upper bounds on
$\norm{\Pi^{\brho^{I_1,I_2}}_\tau (\bb \otimes \bb)}^2$.
\begin{lemma}
\label{lem:mubbound}
Fix $I \subseteq [k]$, $I \neq \{\}$. Then,
$\displaystyle
\E_{\brho,\bb,g}[|\qform{\bb}{\brho^I(g)}{\bb}|^2] 
\leq \sum_{\tau \in \hG} \frac{d_\tau}{|G|}.
$
\end{lemma}
\begin{proof}
We use the notation $\tau \prec \brho^I$ to denote a single copy
of $\tau \in \hG$ occurring in the Clebsch-Gordan decomposition of
$\brho^I$ i.e. treating $\brho^I$ as a representation of $G$ embedded 
in the diagonal of $G^k$. A given $\tau \in \hG$ can occur more than
once in the decomposition, or not at all. We let $\bb_\tau$ denote
the orthogonal projection of $\bb$ onto this copy of $\tau$. Note
that if $\tau$ occurs more than once, then there will be several
orthogonal vectors $\bb_\tau$. If $\norm{\bb_\tau} > 0$, define 
$\hbb_\tau$ to be $\bb_\tau$ normalized; otherwise, let
$\hbb_\tau$ be an arbitrary unit vector in the copy of $\tau$ under
consideration. We now have
\begin{eqnarray*}
&      &
\!\!\!\!\!
\!\!\!\!\!
\!\!
|\qform{\bb}{\brho^I(g)}{\bb}|^2 
    =    \left|\qform{\bb}{\bigoplus_{\tau \prec \brho^I} \tau(g)}
                {\bb}
         \right|^2 
    =    \left|\sum_{\tau \prec \brho^I} 
               \qform{\bb_\tau}{\tau(g)}{\bb_\tau}
         \right|^2 \\
&   =  & \left|\sum_{\tau \prec \brho^I} 
               \norm{\bb_\tau} \cdot \norm{\bb_\tau}
	       \qform{\hbb_\tau}{\tau(g)}{\hbb_\tau}
         \right|^2 \\
& \leq & \left(\sum_{\tau \prec \brho^I} \norm{\bb_\tau}^2\right)
         \cdot 
         \left(\sum_{\tau \prec \brho^I} \norm{\bb_\tau}^2 
               |\qform{\hbb_\tau}{\tau(g)}{\hbb_\tau}|^2 
         \right) \\
&   =  & \sum_{\tau \prec \brho^I} \norm{\bb_\tau}^2 
         \left|\qform{\hbb_\tau}{\tau(g)}{\hbb_\tau}\right|^2.
\end{eqnarray*}
The inequality above follows from Cauchy-Schwartz, and
the last equality is because 
$\sum_{\tau \prec \brho^I} \norm{\bb_\tau}^2 = \norm{\bb}^2 = 1$.
Now,
\begin{eqnarray*}
\lefteqn{\E_{\brho,\bb,g}[|\qform{\bb}{\brho^I(g)}{\bb}|^2]} \\
& \leq & \E_{\brho,\bb,g}
         \left[\sum_{\tau \prec \brho^I} \norm{\bb_\tau}^2 
              \left|\qform{\hbb_\tau}{\tau(g)}{\hbb_\tau}\right|^2 
         \right] \\
&   =  & \E_{\brho,\bb}
         \left[\sum_{\tau \prec \brho^I} \norm{\bb_\tau}^2 
               \E_g\left[\left|\qform{\hbb_\tau}{\tau(g)}{\hbb_\tau}
                         \right|^2
                   \right]
         \right] \\
&   =  & \E_{\brho,\bb}
         \left[\sum_{\tau \prec \brho^I} 
               \frac{\norm{\bb_\tau}^2}{d_\tau} 
         \right] 
    =    \E_{\brho}
         \left[\sum_{\tau \prec \brho^I} 
               \frac{\E_{\bb}[\norm{\bb_\tau}^2]}{d_\tau} 
         \right] \\ 
&   =  & \E_{\brho}
         \left[\sum_{\tau \prec \brho^I} 
               \frac{d_\tau}{d_\tau d_{\brho}} 
         \right]  
    =    \E_{\brho}
         \left[\sum_{\tau \in \hG} 
               \frac{a^{\brho^I}_\tau}{d_{\brho}} 
         \right] 
    =    \sum_{\tau \in \hG} 
               \E_{\brho}\left[\frac{a^{\brho^I}_\tau}{d_{\brho}} 
                         \right] \\
&   =  & \sum_{\tau \in \hG} \frac{d_\tau}{|G|}.
\end{eqnarray*}
The second equality follows from Fact~\ref{likemub}, the fourth
equality follows from Fact~\ref{expected-projection-length} and the
last equality follows from Fact~\ref{expected-multiplicity}.
\end{proof}

The next lemma ties up the above threads to prove an 
upper bound on the second moment of the function $X$ independent
of $k$.
\begin{lemma}
\label{lem:delta1}
$\displaystyle
\E_{\brho,\bb,g}[X(\brho,\bb,g)^2] < 
\ee + \frac{1}{|G|} \cdot
\left(\sum_{\nu \in \hG} d_\nu\right) \cdot
\left(\sum_{\tau \in \cS_\ee} d_\tau |\chi_\tau(h)| \right).
$
\end{lemma}
\begin{proof}
First, note that 
\begin{eqnarray*}
\lefteqn{\E_g[X(\brho,\bb,g)^2]} \\
&   =  & \left|\frac{1}{4^k} \sum_{I_1, I_2 \neq \{\}}
               \sum_{\tau \in \hG} \frac{\chi_\tau(h)}{d_\tau}
               \norm{\Pi^{\brho^{I_1,I_2}}_\tau (\bb \otimes \bb)}^2
         \right| \\
& \leq & \frac{1}{4^k} \sum_{I_1, I_2 \neq \{\}}
         \sum_{\tau \in \hG} \frac{|\chi_\tau(h)|}{d_\tau}
         \norm{\Pi^{\brho^{I_1,I_2}}_\tau (\bb \otimes \bb)}^2 \\
&   <  & \frac{1}{4^k} \sum_{I_1, I_2 \neq \{\}}
         \left( 
         \ee \cdot \sum_{\tau \in \hG \setminus \cS_\ee} 
         \norm{\Pi^{\brho^{I_1,I_2}}_\tau (\bb \otimes \bb)}^2 + 
         \sum_{\tau \in \cS_\ee} \frac{|\chi_\tau(h)|}{d_\tau}
         \norm{\Pi^{\brho^{I_1,I_2}}_\tau (\bb \otimes \bb)}^2 
         \right) \\
&   <  & \ee + \frac{1}{4^k} \sum_{I_1, I_2 \neq \{\}}
         \sum_{\tau \in \cS_\ee} \frac{|\chi_\tau(h)|}{d_\tau}
         \norm{\Pi^{\brho^{I_1,I_2}}_\tau (\bb \otimes \bb)}^2. 
\end{eqnarray*}
The equality follows from Lemma~\ref{lem:chioverd} and the
fact that the quantity in the absolute value sign is non-negative,
and the last inequality follows from the fact that
$\sum_{\tau \in \hG \setminus \cS_\ee}
 \norm{\Pi^{\brho^{I_1,I_2}}_\tau (\bb \otimes \bb)}^2 \leq
 \norm{\bb \otimes \bb}^2 = 1$.

Fix $I_1,I_2 \subseteq [k]$, $I_1,I_2 \neq \{\}$. Then,
\begin{eqnarray*} 
\lefteqn{
\E_{\brho,\bb} 
\left[
\sum_{\tau \in \cS_\ee} \frac{|\chi_\tau(h)|}{d_\tau}
\norm{\Pi^{\brho^{I_1,I_2}}_\tau (\bb \otimes \bb)}^2
\right]
} \\
& \leq &
\E_{\brho,\bb}
\left[
\sum_{\tau \in \cS_\ee} \frac{|\chi_\tau(h)|}{d_\tau} \cdot
\frac{d_\tau^2}{2} 
\left(
\E_g[|\qform{\bb}{\brho^{I_1}(g)}{\bb}|^2] +
\E_g[|\qform{\bb}{\brho^{I_2}(g)}{\bb}|^2]
\right)
\right] \\
&   =  &
\left(
\sum_{\tau \in \cS_\ee} \frac{d_\tau |\chi_\tau(h)|}{2}
\right)
\left(
\E_{\brho,\bb,g}[|\qform{\bb}{\brho^{I_1}(g)}{\bb}|^2] +
\E_{\brho,\bb,g}[|\qform{\bb}{\brho^{I_2}(g)}{\bb}|^2]
\right) \\
& \leq &
\frac{1}{|G|} \cdot
\left(\sum_{\tau \in \cS_\ee} d_\tau |\chi_\tau(h)| \right) \cdot
\left(\sum_{\nu \in \hG} d_\nu\right).
\end{eqnarray*}
The first inequality is due to Lemma \ref{lem:proj-hom} and the second
inequality is due to Lemma \ref{lem:mubbound}. Combining the above
two upper bounds proves the present lemma.
\end{proof}

We now connect the function $X$ to the total variation distance 
between $\cM_{H^g}$ and $\cM_{\{1\}}$.
\begin{lemma}
\label{lem:Xtotvar}
Define $\mu_g := \E_{\brho,\bb}[|X(\brho,\bb,g)|]$. Suppose
$2 k \ee < 1$. Then,
\[
\totvar{\cM_{H^g} - \cM_{\{1\}}} 
< 2^k(1 + 2 k \ee) \mu_g + 3 k \ee + 
  \frac{3 k}{|G|} \cdot \sum_{\tau \in \cS_\ee} d_\tau^2.
\]
\end{lemma}
\begin{proof}
If the hidden subgroup is $H^g$ for some $g \in G$, the probability
of observing an irrep $\brho^\ast \in \hG^{\otimes k}$, row index
$\bi \in [d_{\brho}]$ and 
frame vector $\bb \in \bcB$ is given by 
\[
\cM_{H^g}(\brho,\bi,\bb) = 
\cM_H(\brho) \cdot \frac{1}{d_{\brho}} \cdot
\cM_{H^g}(\bb \mid \brho^\ast).
\]
If the hidden subgroup is $\{1\}$, the probability
of observing an irrep $\brho^\ast \in \hG^{\otimes k}$, row index
$\bi \in [d_{\brho}]$ and 
frame vector $\bb \in \bcB$ is given by 
\[
\cM_{\{1\}}(\brho,\bi,\bb) = 
\cP(\brho) \cdot \frac{1}{d_{\brho}} \cdot
\cN(b \mid \brho).
\]
Define a new probability vector $\cM'_{H^g}$ as 
\[
\cM'_{H^g}(\brho,\bi,\bb) := 
\cP(\brho) \cdot \frac{1}{d_{\brho}} \cdot
\cM_{H^g}(b \mid \brho^\ast).
\]
Define a set 
$\bcS_\ee := 
 \{\brho \in \hG^{\otimes k}: \exists i \in [k], 
                              \rho_i \in \cS_\ee\}$.
Define another new vector $\cM''_{\{1\}}$ with non-negative 
entries as 
\[
\cM''_{\{1\}}(\brho,\bi,\bb) := 
\left\{
\begin{array}{l l}
\cP(\brho) \cdot \frac{1}{d_{\brho}} 
\cdot\frac{a_{\bb}}{2^k r_{\brho}(H^k)}, 
& \mbox{{\rm if}} ~ \brho \not \in \bcS_\ee, \\
0
& \mbox{{\rm otherwise}}
\end{array}
\right..
\]
Note that $\cM''_{\{1\}}$ may not be a probability vector.

Define $D_\ee := \sum_{\tau \in \cS_\ee} d_\tau^2$.
Let $\cP(\cS_\ee)$, $\cP(\bcS_\ee)$ denote the probabilities of 
$\cS_\ee$, $\bcS_\ee$ under
the Plancherel distributions on $\hG$, $\hG^{\otimes k}$ respectively.
Then,
$\cP(\bcS_\ee) \leq k \cP(\cS_\ee) = \frac{k D_\ee}{|G|}$.
Also since 
\[
\frac{d_{\brho}}{2^k r_{\brho}(H^k)} 
= \prod_{i=1}^k \frac{d_{\rho_i}}{2 r_{\rho_i}}
= \prod_{i=1}^k \frac{d_{\rho_i}}{d_{\rho_i} + \chi_{\rho_i}(h)}
= \prod_{i=1}^k \left(1 + \frac{\chi_{\rho_i}(h)}{d_{\rho_i}}
                \right)^{-1},
\]
we have
\[
(1+\ee)^{-k} \leq \frac{d_{\brho}}{2^k r_{\brho}(H^k)} \leq
(1-\ee)^{-k},
\quad \mbox{{\rm for}} \; \brho\in \hG^{\otimes k}\setminus \bcS_\ee.
\]
By the convexity of the function $y = x^{-k}$, we have that
$|(1-\ee)^{-k} - 1| \geq |(1+\ee)^{-k} - 1|$.  Since $2 k \ee < 1$, it
can be shown by induction that $(1-\ee)^{-k} \leq 1 + 2 k \ee$.
Hence,
\[
\left|1 - \frac{d_{\brho}}{2^k r_{\brho}(H^k)}\right| 
\leq 2 k \ee, 
\quad \mbox{{\rm for}} \; \brho\in \hG^{\otimes k}\setminus \bcS_\ee.
\]
Now,
\begin{eqnarray*}
\lefteqn{\totvar{\cM''_{\{1\}} - \cM_{\{1\}}}} \\
&   =  &
\sum_{\brho \in \hG^{\otimes k} \setminus \bcS_\ee} 
\sum_{\bi=1}^{d_{\brho}} \sum_{\bb \in \bcB}
\left|
\cP(\brho) \cdot \frac{1}{d_{\brho}} \cdot 
\frac{a_{\bb}}{2^k r_{\brho}(H^k)} -
\cP(\brho) \cdot \frac{1}{d_{\brho}} \cdot 
\frac{a_{\bb}}{d_{\brho}}
\right| + 
\sum_{\brho \in \bcS_\ee} 
\sum_{\bi=1}^{d_{\brho}} \sum_{\bb \in \bcB}
\cP(\brho) \cdot \frac{1}{d_{\brho}} \cdot 
\frac{a_{\bb}}{d_{\brho}} \\
&   =  &
\sum_{\brho \in \hG^{\otimes k} \setminus \bcS_\ee} 
\sum_{\bb \in \bcB}
\cP(\brho) \cdot \frac{a_{\bb}}{d_{\brho}}
\left|\frac{d_{\brho}}{2^k r_{\brho}(H^k)} - 1 \right| + 
\sum_{\brho \in \bcS_\ee} \cP(\brho) \\
& \leq  &
\sum_{\brho \in \hG^{\otimes k} \setminus \bcS_\ee} 
\cP(\brho) \cdot 2 k \ee + \frac{k D_\ee}{|G|} \\
& \leq &
2 k \ee + \frac{k D_\ee}{|G|}.
\end{eqnarray*}
Next,
\begin{eqnarray*}
\lefteqn{\totvar{\cM'_{H^g} - \cM''_{\{1\}}}} \\
&   =  &
\sum_{\brho \in \hG^{\otimes k} \setminus \bcS_\ee} 
\sum_{\bi=1}^{d_{\brho}} \sum_{\bb \in \bcB}
\left|
\cP(\brho) \cdot \frac{1}{d_{\brho}} \cdot 
\cM_{H^g}(\bb \mid \rho^\ast) -
\cP(\brho) \cdot \frac{1}{d_{\brho}} \cdot 
\frac{a_{\bb}}{2^k r_{\brho}(H^k)}
\right| \\
&      &
+ \sum_{\brho \in \bcS_\ee} 
  \sum_{\bi=1}^{d_{\brho}} \sum_{\bb \in \bcB}
  \cP(\brho) \cdot \frac{1}{d_{\brho}} \cdot 
  \cM_{H^g}(\bb \mid \rho^\ast) \\
&   =  &
\sum_{\brho \in \hG^{\otimes k} \setminus \bcS_\ee} 
\cP(\brho)
\sum_{\bb \in \bcB} 
\left|
\frac{a_{\bb} \qform{\bb}{\brho((H^g)^k)}{\bb}}{r_{\brho}(H^k)} - 
\frac{a_{\bb}}{2^k r_{\brho}(H^k)}
\right| + \sum_{\brho \in \bcS_\ee} \cP(\brho) \\
& \leq &
\sum_{\brho \in \hG^{\otimes k} \setminus \bcS_\ee} 
\cP(\brho) \cdot \frac{d_{\brho}}{r_{\brho}(H^k)}
\sum_{\bb \in \bcB} \frac{a_{\bb}}{d_{\brho}}
\left|
\qform{\bb}{\brho((H^g)^k)}{\bb} - \frac{1}{2^k} 
\right| + \frac{k D_\ee}{|G|} \\
& \leq &
2^k (1 + 2 k \ee)
\sum_{\brho \in \hG^{\otimes k} \setminus \bcS_\ee} \cP(\brho) \cdot
\E_{\bb}[|X(\brho,\bb,g)|] + \frac{k D_\ee}{|G|} \\
& \leq &
2^k (1 + 2 k \ee)
\E_{\brho,\bb}[|X(\brho,\bb,g)|] + \frac{k D_\ee}{|G|} 
    =   
2^k (1 + 2 k \ee) \mu_g + \frac{k D_\ee}{|G|}.
\end{eqnarray*}
Furthermore,
\begin{eqnarray*}
&      &
\!\!\!\!\!
\!\!\!\!\!
\!\!
\totvar{\cM'_{H^g} - \cM_{H^g}} 
    =   
\sum_{\brho \in \hG^{\otimes k}} \sum_{\bi=1}^{d_{\brho}} 
\sum_{\bb \in \bcB}
\left|
\cP(\brho) \cdot \frac{1}{d_{\brho}} \cdot 
\cM_{H^g}(\bb \mid \rho^\ast) -
\cM_{H}(\brho) \cdot \frac{1}{d_{\brho}} \cdot 
\cM_{H^g}(\bb \mid \rho^\ast)
\right| \\
&   =  &
\sum_{\brho \in \hG^{\otimes k}} 
|\cP(\brho) - \cM_{H}(\brho)| 
  \leq  
k \cdot 
\sum_{\tau \in \hG} 
\left| 
\frac{d_\tau^2}{|G|} - \frac{d_\tau |H| r_\tau(H)}{|G|}
\right| 
    =   
k \cdot 
\sum_{\tau \in \hG} 
\left| 
\frac{d_\tau^2}{|G|} - \frac{d_\tau (d_\tau + \chi_\tau(h))}{|G|}
\right| \\
&   =  &
k \cdot 
\sum_{\tau \in \hG} 
\frac{d_\tau |\chi_\tau(h)|}{|G|} 
  \leq  
k \cdot 
\left(
\sum_{\tau \in \hG \setminus \cS_\ee} 
\frac{d_\tau^2}{|G|} \cdot \frac{|\chi_\tau(h)|}{d_\tau} +
\sum_{\tau \in \cS_\ee} 
\frac{d_\tau^2}{|G|}
\right) 
    <   
k \cdot 
\left(
\ee \sum_{\tau \in \hG \setminus \cS_\ee} 
\frac{d_\tau^2}{|G|} +
\frac{D_\ee}{|G|}
\right) \\
& \leq &
k \ee + \frac{k D_\ee}{|G|}.
\end{eqnarray*}
The first inequality follows from $k$ applications of the triangle
inequality. Finally, 
\begin{eqnarray*}
\totvar{\cM_{H^g} - \cM_{\{1\}}} 
& \leq & 
\totvar{\cM_{H^g} - \cM'_{H^g}} +
\totvar{\cM'_{H^g} - \cM''_{\{1\}}} +
\totvar{\cM''_{\{1\}} - \cM_{\{1\}}} \\
& \leq &
2^k(1 + 2 k \ee) \mu_g + 3 k \ee + 
  \frac{3 k}{|G|} \cdot \sum_{\tau \in \cS_\ee} d_\tau^2.
\end{eqnarray*}
\end{proof}
We are now ready to prove the main theorem of the paper.
\begin{proof}[Proof of Theorem~\ref{mainTheorem}]
The theorem follows from Lemmas~\ref{lem:delta1} and
\ref{lem:Xtotvar}, using the convexity of the square function.
The upper bound on $\delta_1$ follows from
the observation that Cauchy-Schwartz implies that
$
\sum_{\nu \in \hG} d_\nu \leq 
 |\hG|^{1/2} \left(\sum_{\nu \in \hG} d_\nu^2\right)^{1/2} =
 |\hG|^{1/2} |G|^{1/2}.
$
\end{proof}

Finally, we prove a simple lower bound, 
irrespective of the order
of entanglement, on the total number
of coset states $t$ required to distinguish a hidden subgroup
$H^g$ from the identity hidden subgroup. For that, we need the
following theorem.
\begin{theorem}
\label{thm:twooutcome}
Let $G$ be a finite group and $H:=\{1,h\}$ be an order two subgroup of
$G$. Let $t\geq 1$ be an integer. Then,
\[
\trnorm{\E_g\left[\sigma_{H^g}^{\otimes t}\right] - 
         \sigma_{\{1\}}^{\otimes t}} < 
\frac{2^t}{|G|} 
\sum_{\tau \in \hG} d_\tau |\chi_\tau(h)|.
\]
\end{theorem}
\begin{proof}
Let $\brho \in \hG^{\otimes t}$, $I \subseteq [t]$, $I \neq \{\}$.
Using arguments similar to those above, it is easy to see that
\begin{eqnarray*}
&      &
\!\!\!\!\!
\!\!\!\!\!
\!\!
\trnorm{\E_g[2^t \brho((H^g)^t)] - \brho(\{1\}^t)}
    =   
\trnorm{
\E_g\left[\onemat_{d_{\brho}} + \sum_{I \neq \{\}} \brho^I(ghg^{-1})
    \right] - 
\onemat_{d_{\brho}} 
} 
    =    
\trnorm{\sum_{I \neq \{\}} \E_g\left[\brho^I(ghg^{-1})\right]} \\
& \leq &
\sum_{I \neq \{\}} \trnorm{\E_g[\brho^I(ghg^{-1})]} 
    =   
\sum_{I \neq \{\}} 
\trnorm{\bigoplus_{\tau \in \hG} \frac{\chi_\tau(h)}{d_\tau} 
        \bigoplus_{j=1}^{a^{\brho^I}_\tau} \onemat_{d_\tau}} 
    =    
\sum_{I \neq \{\}} 
\sum_{\tau \in \hG} a^{\brho^I}_\tau |\chi_\tau(h)|.
\end{eqnarray*}
Writing the density matrices in the Fourier basis and using
Fact~\ref{expected-multiplicity} we get,
\begin{eqnarray*}
&      &
\!\!\!\!\!
\!\!\!\!\!
\!\!
\trnorm{\E_g\left[\sigma_{H^g}^{\otimes t}\right] - 
                 \sigma_{\{1\}}^{\otimes t}} \\
&   =  &
\trnorm{
\E_g\left[
     \frac{2^t}{|G|^t} \bigoplus_{\brho} 
     \bigoplus_{\bi=1}^{d_{\brho}} 
     \ket{\brho^\ast,\bi}\bra{\brho^\ast,\bi}
     \otimes \brho((H^g)^t)
     \right] - 
\frac{1}{|G|^t} \bigoplus_{\brho} \bigoplus_{\bi=1}^{d_{\brho}}
\ket{\brho^\ast,\bi}\bra{\brho^\ast,\bi} \otimes \brho(\{1\}^t)} \\
&   =  &
\trnorm{
\frac{1}{|G|^t}
\bigoplus_{\brho} \bigoplus_{\bi=1}^{d_{\brho}} 
\ket{\brho^\ast,\bi}\bra{\brho^\ast,\bi} \otimes
(\E_g[2^t\brho((H^g)^t)] - \brho(\{1\}^t))} \\
&   =  &
\frac{1}{|G|^t}
\sum_{\brho} d_{\brho}
\trnorm{\E_g[2^t\brho((H^g)^t)] - \brho(\{1\}^t)} 
  \leq  
\frac{1}{|G|^t}
\sum_{\brho} d_{\brho}
\sum_{I \neq \{\}} 
\sum_{\tau \in \hG} a^{\brho^I}_\tau |\chi_\tau(h)| \\
&   =  &
\sum_{I \neq \{\}} \sum_{\tau \in \hG} |\chi_\tau(h)| 
\left(
\sum_{\brho} \frac{d_{\brho}^2}{|G|^t}
\frac{a^{\brho^I}_\tau}{d_{\brho}} 
\right) 
    =   
\sum_{I \neq \{\}} \sum_{\tau \in \hG} 
\frac{d_\tau |\chi_\tau(h)|}{|G|} \\
&   <  &
\frac{2^t}{|G|} \sum_{\tau \in \hG} d_\tau |\chi_\tau(h)|.
\end{eqnarray*}
\end{proof}

\begin{corollary}
\label{cor:twooutcome}
Any algorithm using a total of $t$ coset states that distinguishes
with constant probability between the case when the hidden
subgroup is trivial and the case when the hidden subgroup is
$H^g$ for some $g \in G$ must satisfy $t = \Omega(\log (1/\eta))$.
\end{corollary}
\begin{proof}
The algorithm can be viewed as a two-outcome POVM that outputs
$1$ with probability at least $2/3$ if the hidden subgroup is
non-trivial, and $0$ with probability at least $2/3$ if the 
hidden subgroup is trivial. Thus, the POVM distinguishes
between the states $\E_g\left[\sigma_{H^g}^{\otimes t}\right]$
and $\sigma_{\{1\}}^{\otimes t}$ with constant total variation
distance. Since the trace distance is always an upper bound on the 
total variation distance, invoking Theorem~\ref{thm:twooutcome}
completes the proof.
\end{proof}

The above corollary shows, for example, that any coset state
based algorithm solving the HSP in $S_n \wr S_2$ needs a total
number of $\Omega(n \log n)$ coset states. In the next section, 
we apply
Theorem~\ref{mainTheorem}  to show a stronger result, namely,
any algorithm solving the HSP in $S_n \wr S_2$
using polynomially many coset states needs to make measurements
entangled across $\Omega(n \log n)$ coset states.
However, Corollary~\ref{cor:twooutcome} can sometimes prove
non-trivial lower bounds on the total number of coset states
for solving the HSP in groups $G$ where Theorem~\ref{mainTheorem} 
can only prove a constant lower bound on the order of entanglement. 
For example, the HSP in groups $G := A \rtimes \ZZ_2$, where
$A$ is an abelian group and $\ZZ_2$ acts on $A$ by inversion can
be solved by an algorithm using a total number of $O(\log |G|)$ coset
states that measures one coset state at a time~\cite{EH:98}.
Using Corollary~\ref{cor:twooutcome}, one can show a matching
$\Omega(\log |G|)$ lower bound on the total number of coset states
when $A$ is the cyclic group
$\ZZ_n$, i.\,e., $G$ is the dihedral group $D_n$. 
Using a different technique,
Childs and Wocjan~\cite{CW:2005} in fact show 
an $\Omega(\log |G|)$ lower bound on the total number of 
coset states for the above groups for all abelian $A$.

\section{Limitations of quantum coset states for HSP: Examples}

\subsection{The wreath product $S_n \wr S_2$ and graph isomorphism}

The representation theory of the wreath product $G = S_n \wr S_2$ is
well-known. The following is a summary of the necessary results, for
more details we refer to Appendix \ref{ap:wreathIrrep}: the wreath
product has irreps $\kappa_{\lambda,\lambda^\prime}$ of dimension
$2d_\lambda d_{\lambda^\prime}$, where 
$\lambda, \lambda^\prime \in \widehat{S_n}$, $\lambda \neq \lambda'$.
Define $h := (e,e,1) \in G$, where $e$ is the identity permutation
in $S_n$.
The character value of $h$ on these irreps is zero.
Furthermore, there are irreps $\vartheta_\lambda$ and
$\vartheta_\lambda'$ of dimension $d_\lambda^2$, where $\lambda \in
\widehat{S_n}$. The character values of $\vartheta_\lambda$ and
$\vartheta_\lambda'$ on $h$ are given by $d_\lambda$ and $-d_\lambda$,
respectively. The total number of irreps of $G$ is
$|\hG| = {p(n) \choose 2} + 2 p(n) \leq p(n)^2$, where $p(n)$ denotes
the number of partitions of $n$.

In order to apply Theorem \ref{mainTheorem} we choose $\ee =
n^{-\alpha n}$ for some constant $\alpha > 0$ to be determined later.
Then
$
\cS_\ee = \left\{\sigma\in \hG :
\frac{|\chi_\sigma(h)|}{d_\sigma} \geq \ee\} =
\{\vartheta_\lambda,\vartheta_\lambda': d_\lambda 
\leq n^{\alpha n}\right\}. 
$
Hence we obtain that
\[
\sum_{\sigma\in \cS_\ee} d_\sigma \cdot|\chi_\sigma(h)| \leq 2
\sum_{\lambda\in \widehat{S_n}, d_\lambda \leq n^{\alpha n}} 
d_\lambda^2
\cdot d_\lambda \leq p(n) n^{2\alpha n} \cdot n^{\alpha n} \leq n^{3
  \alpha n} e^{\nu\sqrt{n}}.
\]
Here we have estimated the partition number as $p(n) = O(e^{\nu
  \sqrt{n}})$, where $\nu = \pi \sqrt{\frac{2}{3}}$. We also compute
that
\[
\sum_{\sigma\in \cS_\ee} d_\sigma^2 \leq 2
\sum_{\lambda\in \widehat{S_n}, d_\lambda \leq n^{\alpha n}} 
d_\lambda^4
\leq p(n) n^{4\alpha n} \leq n^{4 \alpha n} e^{\nu\sqrt{n}}.
\]
In order to apply Theorem~\ref{mainTheorem}, we now define 
$\alpha := 1/4$ and obtain that
\begin{eqnarray*}
\delta_1 & \leq &
\ee + \left(\sum_{\sigma \in\cS_\ee}
    d_\sigma|\chi_\sigma(h)| \right) 
    \left(\frac{|\hG|}{|G|}\right)^{1/2}
\leq
n^{-\alpha n} + n^{3\alpha n} e^{\nu \sqrt{n}}
    \left({\frac{p(n)^2}{2 (n!)^2}}\right)^{1/2}\\
& \leq &
n^{-1/4 n} + \frac{n^{3/4 n} e^{2 \nu \sqrt{n}}}{\sqrt{2} n!}
= 
n^{-\Omega(n)},
\end{eqnarray*}
where we have used the fact that $n! \geq (n/e)^n$ for large $n$.
For the parameter
$\delta_2$ in Theorem~\ref{mainTheorem} we obtain
\begin{align*}
\delta_2  & = 2^k(1+2k\ee)\delta_1^{1/2} + 3k\ee + 
\frac{3k\sum_{\sigma \in\bcS_\ee} d_\sigma^2}{|G|} \\
& \leq 2^k
\left(1 + 2 k n^{-1/4 n} \right)
n^{-\Omega(n)} + 3 k n^{-1/4 n} + 3k \frac{n^{n} e^{\nu\sqrt{n}}}
{2(n !)^2} = 2^k n^{-\Omega(n)}.
\end{align*}
Hence, we have proved the following corollary to Theorem
\ref{mainTheorem}:
\begin{corollary}
\label{cor:wreath}
Any algorithm operating on coset states that solves the hidden
subgroup problem in $G = S_n \wr S_2$ in polynomial time has to make
joint measurements on $k = \Omega(n \log n)$ coset states. The
same is true for any algorithm that solves the
hidden subgroup problem in $S_n$ using coset states. 
Also, any efficient
algorithm for isomorphism of two $n$-vertex graphs that uses 
the standard reduction to HSP in $S_{2n}$ and then uses coset states
to solve the HSP needs to make measurements entangled across
$k = \Omega(n \log n)$ coset states.
\end{corollary}

Finally, we remark that if we apply Theorem~\ref{mainTheorem}
to all the full-support involutions in $S_{2n}$, we only get
a lower bound of $k = \Omega(n)$. This is because we use
Roichman's~\cite{Roichman:96} upper bound on the normalized
characters of $S_{2n}$ in order to define $\cS_\epsilon$, as
in \cite{MooreRS05}, and Roichman's bound is always at least
$e^{-O(n)}$. Since the involutive swaps form an exponentially
small fraction of all the full-support involutions, it is 
possible that an average hidden full-support involution may be 
distinguishable from the hidden identity subgroup by 
an $O(n)$-entangled POVM acting on $n^{O(1)}$-coset states.
However, no such POVM is known and the best upper bound for
this problem continues to be the $k = O(n \log n)$ 
information-theoretic one.

\subsection{The projective linear groups $\PSL(2,\FF_q)$}

The representation theory of the projective linear groups $G =
\PSL(2,\FF_q)$ over any finite field $\FF_q$ is well-known.  The
following is a summary of the necessary results, for more details we
refer to Appendix \ref{ap:projIrrep}. We treat the cases $q$ even and
$q$ odd separately. In case $q$ odd we have that $|\PSL(2,\FF_q)| =
\frac{q(q^2-1)}{2}$. There is one conjugacy class of $\frac{q(q\pm
  1)}{2}$ involutions (depending on whether $q \equiv 1$ or $3$ modulo
$4$); let $h$ denote a fixed member of this conjugacy class.
The degrees of the irreps are given by $1, q, q\pm 1$, and
$\frac{q\pm 1}{2}$. The character values 
$|\chi(h)|$ can be upper bounded by $1$, $1$, $2$, and
$1$, respectively. There is a total number of $|\hG| =
\frac{q+5}{2}$ irreps.

In order to apply Theorem \ref{mainTheorem}, we choose $\ee =
\frac{2}{q-1}$. Then 
\[
\cS_\ee 
= \{\sigma\in \hG: \frac{|\chi_\sigma(h)|}{d_\sigma} \geq \ee\} 
= \{\onemat \}
\]
contains only the trivial irrep. With this choice of the parameter
$\ee$ we have that
\[
\sum_{\sigma\in \bcS_\ee} d_\sigma \cdot|\chi_\sigma(h)| = 1, \quad
\sum_{\sigma\in \bcS_\ee} d_\sigma^2 = 1, \quad \text{and} \quad
\left(\frac{|\hG|}{|G|}\right)^{1/2} = \left(\frac{(q+5)/2}{q
    (q^2-1)/2}\right)^{1/2} = O(q^{-1}).
\]
Hence, we can bound the parameter $\delta_1$ used in
Theorem~\ref{mainTheorem} as follows:
\[
\delta_1  \leq
\ee + \left(\sum_{\sigma \in\bcS_\ee}
    d_\sigma|\chi_\sigma(h)| \right) 
    \left(\frac{|\hG|}{|G|}\right)^{1/2}
 \leq  
\frac{2}{q-1} + 1 \cdot O(q^{-1})
= O(q^{-1}).
\]
For the parameter $\delta_2$ we obtain
\begin{align*}
\delta_2  & = 2^k(1+2k\ee)\delta_1^{1/2} + 3k\ee + 
\frac{3k\sum_{\sigma \in\bcS_\ee} d_\sigma^2}{|G|} \\ 
& \leq 2^k \left(1+2k \frac{2}{q-1}\right) O(q^{-1/2}) 
+ 3k \frac{2}{q-1} + 
3k \frac{1}{q(q^2-1)/2} \leq 2^k O(q^{-1/2}).
\end{align*}
The case $q=2^n$, where $|\PSL(2,\FF_{2^n})| =
|\SL(2,\FF_{2^n})| =q(q^2-1)$, can be treated similarly. There we use
$\ee = \frac{1}{q-1}$ which implies that $\delta_2 \leq 2^k
O(q^{-1/2})$. Hence, using Theorem \ref{mainTheorem} we have shown the
following result:
\begin{corollary}\label{cor:PSL}
  Let $q$ be a prime power. Then any algorithm operating on coset
  states that solves the hidden subgroup problem in $G =
  \PSL(2,\FF_q)$ in polynomial time has to make joint measurements on
  $k = \Omega(\log |G|) = \Omega(q)$ coset states. 
\end{corollary}

\subsection{Special and general linear groups}

\begin{corollary}\label{cor:GL}
  Any algorithm solving the HSP in $\SL(2,\FF_q)$ or $\GL(2,\FF_q)$
  efficiently using coset states needs to make measurements entangled
  across $k=\Omega(\log{q})$ registers.
\end{corollary}
\begin{proof}
  By Corollary \ref{cor:PSL} any algorithm solving the HSP in
  $\PSL(2,\FF_q)$ efficiently using coset states needs to make
  measurements entangled across $k = \Omega(\log q)$ registers. The
  statement now follows from Lemma~\ref{lem:transfer} by using the
  facts that $\PSL(2,\FF_q) \cong \SL(2,\FF_q)/\zeta(\SL(2,\FF_q))$
  and that $\SL(2,\FF_q) \leq \GL(2,\FF_q)$.
\end{proof}

\begin{corollary}
Any algorithm solving the HSP in $\GL(n,\FF_{p^m})$ efficiently
using coset states needs to make measurements entangled across
$k=\Omega(n (m \log p + \log n))$ registers.
\end{corollary}
\begin{proof}
Since $\GL(n,\FF_{p^m})$ contains all $n \times n$ permutation
matrices,
a lower bound of $k = \Omega(n \log n)$ follows from
Corollary~\ref{cor:wreath} and Lemma~\ref{lem:transfer}.
Also, we can use 
the embedding of $\GL(2,\FF_{p^{nm}}) \leq \GL(2n,\FF_{p^m})$
via 
$
\left(\begin{array}{rr} a & b \\ c & d \end{array} \right)
\mapsto 
\left(\begin{array}{rr} M_a & M_b \\ M_c & M_d \end{array} \right),
$
where for each $x \in \FF_{p^{nm}}$ the matrix $M_x \in
\GL(n,\FF_{p^m})$ realizes multiplication by
$x$ with respect to a fixed basis of 
$\FF_{p^{nm}}$ over $\FF_{p^m}$.  Hence by 
Lemma~\ref{lem:transfer} we
obtain that for the HSP in $\GL(2n,\FF_{p^m})$ at least as much
entanglement is necessary as in case of $\GL(2,\FF_{p^{nm}})$. The
latter has been bounded by $\Omega(nm \log p)$ in 
Corollary \ref{cor:GL}.
\end{proof}

\subsection{Direct products of the form $G^n$}
 
In this section we show that for a large class of finite groups $G$,
efficient algorithms for 
HSP for direct products of the form $G^n$, where $n \geq 1$,
require entangled measurements on at least $k = \Omega(n)$ coset
states. 
Let $G$ be a finite group and let $\hG = \{\sigma_1,
\ldots, \sigma_m\}$ denote the irreducible representations of $G$.
Recall that the centralizer $C(g)$ of an element $g \in G$ is the
subgroup $C(g) := \{ c \in G: cg=gc\}$. Let $h$ be an involution in
$G$, and let $\sigma\in \hG$. Then either
$|\chi_\sigma(h)|=d_\sigma$ or $\frac{|\chi_\sigma(h)|}{d_\sigma}<
1-\frac{2|C(h)|}{|G|}$ holds \cite{Gallagher:94}.  We define $\ee :=
(1-\frac{2|C(h)|}{|G|})^t$, where $t=t(n)$ is a function of $n$ to be
determined later.

The irreps of $G^n$, where $n \geq 1$, are given by 
$\bsigma := \sigma_1 \otimes \ldots \otimes \sigma_n$, where 
$\sigma_i \in \hG$.  
We let $\Lambda := \{ \sigma \in \hG : |\chi_\sigma(h)| =
d_\sigma\}$, $\lambda := \sum_{\sigma \in \Lambda} d_\sigma^2$, and
$\mu := \sum_{\sigma \in \hG \setminus \Lambda} d_\sigma^2
        = |G| - \lambda$.  
The following property of the set 
\[
\bcS_\ee := \left\{\bsigma \in \hG^n :
\frac{|\chi_{\bsigma}(h, \ldots, h)|}{d_\sigma} \geq \ee\right\}
\]
holds for our choice of
the parameter $\ee$: if $\bsigma \in \bcS_\ee$ then 
necessarily at least
$n-t$ positions $\sigma_i$ have to be from $\Lambda$, i.\,e., have to
satisfy $|\chi_{\sigma_i}(h)|=d_{\sigma_i}$. Indeed, otherwise we
would have more than $t$ positions $\sigma_j$ in each of which
$\frac{|\chi_{\sigma_j}(h)|}{d_{\sigma_j}} \leq 
1-\frac{2|C(h)|}{|G|}$, making the product less than $\ee$. 
We next give an estimate for the quantity 
$\sum_{\bsigma \in \bcS_\ee} d_{\bsigma}^2$ appearing in 
Theorem~\ref{mainTheorem}.
For that we require the following lemma 
for estimating the tail of the binomial distribution.
\begin{lemma} 
\label{lemma:binomialbound}
Let $\alpha, \beta >0$, let $n \geq 1$, and let $t = n/c$, 
where
$c > \frac{\alpha+\beta}{\beta}$. Then 
\[
\sum_{\ell = n-t}^n {n \choose \ell} \alpha^\ell \beta^{n-\ell}
\leq \left(\alpha 
\left(\frac{c e (\alpha+\beta)}{\alpha}\right)^{1/c}\right)^n.
\]
\end{lemma}
\begin{proof}
We have that 
\begin{eqnarray*}
&      &
\!\!\!\!\!
\!\!\!\!\!
\!\!
\sum_{\ell = n-t}^n {n \choose \ell} \alpha^\ell \beta^{n-\ell}
    =
(\alpha+\beta)^n \sum_{\ell = n-t}^n {n \choose \ell} 
\left(\frac{\alpha}{\alpha+\beta}\right)^\ell 
\left(\frac{\beta}{\alpha+\beta}\right)^{n-\ell} \\
& \leq  &
(\alpha+\beta)^n {n \choose n-t} 
\left(\frac{\alpha}{\alpha+\beta}\right)^{n-t} \\
&   =  & \alpha^n {n \choose t} 
         \left(\frac{\alpha+\beta}{\alpha}\right)^t 
  \leq   \alpha^n \left(\frac{ne (\alpha+\beta)}{t\alpha}\right)^t 
    =    \left(\alpha 
         \left(\frac{c e (\alpha+\beta)}{\alpha}\right)^{1/c}
         \right)^n,
\end{eqnarray*}
where the first inequality follows from the union bound on 
probabilities and the second one from 
${n \choose t} \leq \left(
  \frac{ne}{t}\right)^t$.
\end{proof}

Suppose we fix $\ell \geq n-t$
locations for putting in irreps from $\Lambda$.
The contribution of this configuration 
to $\sum_{\bsigma \in \bcS_\ee} d_{\bsigma}^2$ 
is the sum of products
of squares of dimensions of $\ell$ irreps from $\Lambda$ and
$n - \ell$ irreps from $\hG \setminus \Lambda$, which simplifies to
$\lambda^{\ell} \mu^{n-\ell}$. 
Letting
$\alpha := \lambda$, $\beta := \mu$, and $t = n/c$, with
some constant $c$ to be determined later, we obtain the following
bound from Lemma~\ref{lemma:binomialbound}:
\[
\sum_{\bsigma \in \bcS_\ee} d_{\bsigma}^2 \leq
\sum_{\ell = n-t}^n {n \choose \ell} \lambda^\ell \mu^{n-\ell} 
\leq \lambda^n \left(\left(\frac{c e |G|}{\lambda}
                     \right)^{1/c}\right)^n.
\]
Hence, for any given $\kappa > 0$ we can find a constant $c>0$ 
such that $\sum_{\bsigma \in \bcS_\ee} d_{\bsigma}^2
\leq \lambda^n \left(1+\kappa\right)^n$ holds for
all $n \geq c$. Note that the same upper bound applies to
$\sum_{\sigma \in \bcS_\ee} d_{\bsigma} 
                            |\chi_{\bsigma}(h, \ldots, h)|$. 
Also, observe that 
$\sum_{\brho \in \hG^n} d_{\brho} = 
 \left(\sum_{\rho \in \hG} d_\rho\right)^n$.
Now, we can bound the parameter
$\delta_1$ used in Theorem~\ref{mainTheorem}:
\begin{eqnarray*}
\delta_1 
& \leq & \ee + \frac{1}{|G|^n}
         \left(\sum_{\bsigma \in\bcS_\ee} d_{\bsigma} 
              |\chi_{\bsigma}(h, \ldots, h)| 
         \right) 
         \left(\sum_{\brho \in \hG^n} d_{\brho}\right) \\
& \leq & \left(\left(1-\frac{2|C(h)|}{|G|}\right)^{1/c}\right)^n 
         + \frac{\lambda^n (1+\kappa)^n}{|G|^n}
           \left(\sum_{\rho \in \hG} d_\rho\right)^n.
\end{eqnarray*}
For the following we make the assumption that $|G| > \lambda
(1+\kappa) \left(\sum_{\rho \in \hG} d_\rho\right)$ holds. 
This implies that there exists a
constant $\gamma_1>0$ such that $\delta_1 \leq \gamma_1^n$.  For the
parameter $\delta_2$ in Theorem~\ref{mainTheorem} we obtain
\begin{align*}
\delta_2 & = 2^k(1+2k\ee)\delta_1^{1/2} + 3k\ee + 
\frac{3k}{|G|^n} \sum_{\bsigma \in\bcS_\ee} d_{\bsigma}^2 \\
& \leq 2^k
\left(1 + 2 k \left(1-\frac{2|C(h)|}{|G|}\right)^{n/c} 
\right)
\gamma_1^{n/2} + 3 k \left(1-\frac{2|C(h)|}{|G|}\right)^{n/c}
+ 3k \left(\frac{\lambda (1+\kappa)}{|G|}\right)^n.
\end{align*}
Now, since our assumption implies that 
$|G| > \lambda (1 + \kappa)$, we
obtain that there exists a constant $\gamma_2 > 0$ such that $\delta_2
\leq \gamma_2^n$. Hence, we have proved the following 
corollary to Theorem \ref{mainTheorem}.
\begin{corollary}
\label{cor:direct1}
Let $G$ be a finite group and let $h \in G$ be an involution.  Let
$\hG$ denote the set of irreps of $G$ and let $\Lambda := \{
\sigma \in \hG : |\chi_\sigma(h)| = d_\sigma\}$. 
Suppose
that $|G| > \left(\sum_{\sigma \in \Lambda} d_\sigma^2\right)
            \left(\sum_{\rho \in \hG} d_\rho\right)$ holds. Then
any efficient algorithm operating on coset states that
distinguishes between the case when the hidden subgroup is a 
conjugate of the subgroup $\langle (h, \ldots, h) \rangle \leq G^n$,
and the case when the hidden subgroup is the identity subgroup in 
$G^n$, needs to make measurements entangled across  
$\Omega(n)$ registers.
\end{corollary}
Recently, Alagic, Moore and Russell~\cite{AMR:05}
showed that any measurement on a single coset state
gives exponentially little information about a hidden subgroup
in the group $G^n$, where $G$ is fixed and satisfies a suitable 
condition.
Their condition on $G$ is weaker than our condition in 
Corollary~\ref{cor:direct1}, but they
only prove lower bounds for algorithms measuring one coset state
at a time. They also give several examples of families of
groups satisfying their condition, 
including all non-abelian finite simple groups.
In fact, the condition of Corollary~\ref{cor:direct1}
holds for all families of
groups $G$ considered in their paper, showing that efficient
coset state based algorithms solving the HSP for their families
of groups $G^n$ need to make  measurements entangled across
$\Omega(n)$ registers.

From Corollary~\ref{cor:direct1}, it is easy to prove
Corollary~\ref{cor:direct2} via the Cauchy-Schwartz inequality.
\begin{corollary}
\label{cor:direct2}
Let $G$ be a finite group and let $h \in G$ be an involution.  Let
$\hG$ denote the set of irreps of $G$ and let $\Lambda := \{
\sigma \in \hG : |\chi_\sigma(h)| = d_\sigma\}$. 
Suppose that 
$|G|^{1/2} > |\hG|^{1/2} 
             \left(\sum_{\sigma \in \Lambda} d_\sigma^2\right)$ 
holds. Then
any efficient algorithm operating on coset states that
distinguishes between the case when the hidden subgroup is a 
conjugate of the subgroup $\langle (h, \ldots, h) \rangle \leq G^n$,
and the case when the hidden subgroup is the identity subgroup in 
$G^n$, needs to make measurements entangled across  
$\Omega(n)$ registers.
\end{corollary}

Using Corollary~\ref{cor:direct2}, we prove the following result.
\begin{corollary}
\label{cor:sm}
Any efficient algorithm operating on coset states that
distinguishes between the case when the hidden subgroup is a 
conjugate of the subgroup 
$\langle (h, \ldots, h) \rangle \leq (S_m)^n$
where $h\in S_m$ is any involution and $m\geq 5$ is fixed,
and the case when the hidden subgroup is the identity subgroup in 
$(S_m)^n$, needs to make measurements entangled across  
$\Omega(n)$ registers. The same holds also
when $m=4$ and $h=(1,2) \in S_4$. 
\end{corollary}
\begin{proof}
Let $G = S_m$, where $m \geq 5$, and let $h$ be any involution in
$G$. Recall that for $m\geq 5$ all irreps of $S_m$ of degree greater
than $1$ are
faithful \cite[Theorem 2.1.13]{JK:81}, and that the center of $S_m$
is trivial. Since for faithful $\sigma \in \widehat{S_m}$ we have that
$|\chi_\sigma(h)|=d_\sigma$ implies that $h$ is in the center, we
obtain that $|\chi_\sigma(h)|<d_\sigma$ for all 
$\sigma\in\widehat{S_m}$
with $d_\sigma > 1$. Hence $\Lambda = \{ \onemat, {\rm alt} \}$
consists of the trivial and the alternating character only and we
obtain that $\sum_{\sigma \in \Lambda} d_\sigma^2 = 2$. 
Since for $m \geq 5$ we have
that $|G|^{1/2} = \sqrt{m!} > 2 \sqrt{p(m)} = 
|\hG|^{1/2} \sum_{\sigma \in \Lambda} d_\sigma^2$, where 
$p(m)$ denotes the partition number of
$m$, the statement for $m \geq 5$ follows from Corollary
\ref{cor:direct2}.  

For $m=4$ and $h=(1,2)$ we observe that the set
$\Lambda$ is again given by $\Lambda = \{ \onemat, {\rm alt} \}$.
We verify that the condition
$|S_4|^{1/2} = \sqrt{24} > 2 \sqrt{5} = 
|\widehat{S_4}|^{1/2} \sum_{\sigma \in \Lambda} d_\sigma^2$ holds. 
Hence the statement for this case also
follows from Corollary \ref{cor:direct2}.
\end{proof}

\subsection*{Acknowledgments}
We thank Andrew Childs, 
Fr\'{e}d\'{e}ric Magniez and Umesh Vazirani
for helpful discussions and comments.

\bibliographystyle{alpha}
\bibliography{multireg}

\begin{appendix}

\section{Representations of the wreath product $S_n \wr S_2$}
\label{ap:wreathIrrep}

We describe the irreducible representations of the wreath product $S_n
\wr S_2$, i.\,e., the group $(S_n \times S_n) \rtimes Z_2$. We will
also get formulas for the character values under these representations
in terms of the character values of irreducible representations of
$S_n$.

Let $\widehat{S_n} = \{\sigma_i : i = 1, \ldots, p(n)\}$ denote the
irreducible representations of $S_n$, where $p(n)$ denotes the number
of partitions of $n$. Denote the degree of $\sigma_i\in \widehat{S_n}$
by $d_i$.  Letting $N := (S_n \times S_n)$ and $G := (S_n \times S_n)
\rtimes Z_2$ we have that $N \lhd G$ is a normal subgroup of index
$2$. The irreducible representations of $N$ are given by
$\widehat{N} = \{ \sigma_i \otimes \sigma_j : i,j = 1, \ldots, p(n)\}$
and we define the shorthand $\phi_{i,j} := \sigma_i \otimes \sigma_j$.
Define $t := (e,e,1) \in G$, where $e$ is the identity permutation in
$S_n$.  A transversal of $N$ in $G$ is given by $T = \{(e,e,0), t\}$.
Then $t$ acts on $\widehat{N}$ as $(\sigma_i \otimes \sigma_j)^t =
(\sigma_j \otimes \sigma_i)$. Hence we have that $\phi_{i,j}^t =
\phi_{j,i}$. Since all $\phi_{i,j}$ are pair-wise inequivalent, we
obtain the following two cases from Clifford's
Theorem~\cite{Isaacs:76}.

\begin{itemize}
\item[(i)] $i=j$. Then $\phi_{i,j} \cong\phi_{i,j}^t$. Hence
$\phi_{i,j}$ has precisely $2$ pairwise inequivalent extensions to
$G$. One of these extensions is $\vartheta_i = \ext{\phi_{i,i}}$ in
which the image of $t$ permutes the tensor factors of $\CC^{d_i}
\otimes \CC^{d_i}$, where $d_i={\rm deg}(\sigma_i)$. Hence if $\{e_k :
k =1, \ldots, d_i\}$ denotes the standard basis of $\CC^{d_i}$ then
$\vartheta_{i}(t)$ is given by the matrix ${\rm SWAP}_{d_i}$ which
maps $e_k \otimes e_\ell \mapsto e_\ell \otimes e_k$.  The
other extension $\vartheta^\prime_{i}$ of $\phi_{i,i}$ to $G$ is
given by defining the image of $t$ to be $\vartheta^\prime(t) :=
-\vartheta_i(t)$. Note that both extensions have degree $d_i^2$.
The character value ${\rm tr}(\vartheta_i(t))$ is given by the
number of invariant tensors under the swap operation, i.\,e., 
${\rm tr}(\vartheta_i(t)) = d_i$ and
${\rm tr}(\vartheta^\prime_i(t)) = -d_i$.

\item[(ii)] $i \not=j$. Then $\phi_{i,j} \not\cong\phi_{i,j}^t =
\phi_{j,i}$. Hence $\kappa_{i,j} := \phi_{i,j}\ind_T G$ is
irreducible.  Moreover, we have that $(\phi_{i,j}\ind_T G)\res N =
\phi_{i,j} \oplus \phi_{j,i}$ and 
\[
(\phi_{i,j}\ind_T G)(t) =
  \left(\begin{array}{cc}  \zeromat_{d_i d_j} & \onemat_{d_i d_j}\\
      \onemat_{d_i d_j} & \zeromat_{d_i d_j} \end{array} \right).
\]
\end{itemize}

We summarize the facts relevant for this paper in the following table 
by showing the images of elements of the form $(\pi, \mu,
e)$ and $t=(e, e, 1)$ under the irreducible representations of $G
= (S_n \times S_n)\rtimes Z_2$:

\[
\setlength{\arraycolsep}{1.4mm}
\begin{array}{ccccc}
\hline
\mbox{Irrep} & \mbox{Irrep on}\; (\pi, \mu, e) & 
\mbox{Char. on}\; (\pi, \mu, e) & \mbox{Irrep on} \; t & 
\mbox{Char. on} \; t \\
\hline\hline
\vartheta_i & \sigma_i(\pi)\otimes \sigma_i(\mu) & 
\chi_i(\pi) \chi_i(\mu) & 
\phantom{-}{\rm SWAP}_{d_i} & \phantom{-}d_i \\
\vartheta^\prime_i & \sigma_i(\pi)\otimes \sigma_i(\mu) & 
\chi_i(\pi) \chi_i(\mu) & 
-{\rm SWAP}_{d_i} & -d_i \\
\kappa_{i,j} & \left(\begin{array}{cc} 
\sigma_i(\pi)\otimes \sigma_j(\mu) & 
\zeromat_{d_i d_j} \\ \zeromat_{d_i d_j} & \sigma_j(\pi) 
\otimes \sigma_i(\mu) 
\end{array} \right) & 
\chi_i(\pi) \chi_j(\mu) + \chi_j(\pi) \chi_i(\mu) &
\left(\begin{array}{cc} \zeromat_{d_i d_j} & \onemat_{d_i d_j}\\
\onemat_{d_i d_j} & \zeromat_{d_i d_j} \end{array} \right) & 0\\
\hline
\end{array}
\]

Overall, there are ${p(n) \choose 2}$ pairwise inequivalent
irreducible representations $\kappa_{i,j} \in \hG$, one for
each pair $i,j$ such that $i \not= j$. We have that the degree of
$\kappa_{i,j}$ is given by $2 d_i d_j$. The character $\chi_{i,j}$ of
$\kappa_{i,j}$ satisfies $\kappa_{i,j}(t)=0$ for all $i \not= j$.
Furthermore, there are $2 p(n)$ pairwise inequivalent irreducible
representations $\vartheta_i$ and $\vartheta'_i$.

\section{Representations of the projective linear groups 
$\PSL(2,\FF_q)$}
\label{ap:projIrrep}

We briefly recall some facts from the representation theory of the
projective linear groups $\PSL(2,\FF_q)$, where $q$ is a prime power.
Good references on the complex representation theory of these groups
are available, see e\,.g, \cite{BZ:99,FH:91,LR:92}. We treat the cases
$q$ odd and $q=2^n$ separately and begin by describing the conjugacy
classes of involutions and the irreducible representations of
$\PSL(2,\FF_q)$ for $q$ odd. Recall that for $q$ odd, the center of
$\SL(2,\FF_q)$ consists only of the identity matrix and the matrix
\[
c := \left( \begin{array}{rr} -1 & 0 \\ 0 & -1 \end{array}\right).
\]
Once the characters of $\SL(2,\FF_q)$ are known, we therefore have to
filter out only those characters $\chi$ for which $\chi(c) = \chi(1)$
holds in order to obtain the irreducible representations of
$\PSL(2,\FF_q)$.

\subsection{The case $\PSL(2,\FF_q)$ where $q \equiv 1 \mod 4$} The
involutions are given by conjugates of the residue class of
\[
h = \overline{
    \left(
    \begin{array}{rr} 
    0  & 1 \\ 
    -1 & 0
    \end{array}
    \right)} \in \PSL(2,\FF_q),
\]
where the bar denotes the fact that we are using coset
representatives with respect to the center $\langle c \rangle$ of
$\SL(2,\FF_q)$.  There is a total of $\frac{q(q-1)}{2}$ many
involutions that are conjugates of $h$. The characters and their
values on $h$ are summarized in the following table.
\[
\renewcommand{\arraystretch}{1.4}
\begin{array}{ccccc}
\hline
\mbox{Irrep name} & \mbox{Parameters} & \mbox{Number of irreps} &
\mbox{Degree} & \mbox{Character value at}\; h \\
\hline\hline
\onemat & \text{---} & 1 & 1 & 1 \\
\psi & \text{---} & 1 & q & 1 \\
\theta_k & k=2,4,\ldots,\frac{q-1}{2} & \frac{q-1}{4} & q-1 &
0  \\
\chi_j & j=2,4,\ldots,\frac{q-5}{2} & 
\frac{q-5}{4} & q+1 &  2 (-1)^{k/2}\\
\zeta_\ell & \ell = 1, 2 & 2 & \frac{q+1}{2} & (-1)^{(q-1)/4} \\
\hline
\end{array}
\]

\subsection{The case $\PSL(2,\FF_q)$ where $q \equiv 3 \mod 4$}
Similar to the previous case all involutions are conjugate
to the element $h$ defined as above. However, now there are 
$\frac{q(q+1)}{2}$
involutions conjugate to $h$.  The
characters and their values on $h$ are summarized in the following
table.
\[
\renewcommand{\arraystretch}{1.4}
\begin{array}{ccccc}
\hline
\mbox{Irrep name} & \mbox{Parameters} & \mbox{Number of irreps} &
\mbox{Degree} & \mbox{Character value at}\; h \\
\hline\hline
\onemat & \text{---} & 1 & 1 & 1 \\
\psi & \text{---} & 1 & q & -1 \\
\theta_k & k=2,4,\ldots,\frac{q-3}{2} & \frac{q-3}{4} & q-1 &
2 (-1)^{k/2+1} \\
\chi_j & j=2,4,\ldots,\frac{q-3}{2} & \frac{q-3}{4} & q+1 & 0\\
\eta_\ell & \ell = 1, 2 & 2 & \frac{q-1}{2} & 
(-1)^{\frac{q+1}{4}+1} \\
\hline
\end{array}
\]

\subsection{The case $\PSL(2,\FF_q)$ where $q=2^n$}
This case behaves quite differently from the case $q$ odd. First,
observe that in this case the center is trivial, i.\,e.,
$\PSL(2,\FF_{2^n})=\SL(2,\FF_{2^n})$. 
All involutions in $\SL(2,\FF_{2^n})$
are conjugate to the element
\[
h = \left( \begin{array}{rr} 1 & 1 \\ 0 & 1\end{array}\right)
\in \SL(2,\FF_q),
\]
and there is a total number of $q^2-1$ of such involutions.  The
characters and their values on $h$ are summarized in the following
table.
\[
\renewcommand{\arraystretch}{1.4}
\begin{array}{ccccc}
\hline
\mbox{Irrep name} & \mbox{Parameters} & \mbox{Number of irreps} &
\mbox{Degree} & \mbox{Character value on}\; h \\
\hline\hline
\onemat & \text{---} & 1 & 1 & 1 \\
\psi & \text{---} & 1 & q & 0 \\
\theta_k & k=1, 2, \ldots, \frac{q}{2} & \frac{q}{2} & q-1 &
-1 \\
\chi_j & j=1, 2, \ldots, \frac{q-2}{2} & \frac{q-2}{2} & q+1 & 1\\
\hline
\end{array}
\]

\end{appendix}

\end{document}